\documentclass{aastex}
\usepackage[flushleft]{threeparttable}
\usepackage{booktabs,fixltx2e}
\usepackage{graphicx}
\usepackage{amsmath}
\usepackage{subfigure}
\usepackage{amssymb}
\usepackage{tabularx}
\usepackage{natbib}
\usepackage{tikz}
\usetikzlibrary{calc,decorations.markings}

\newcommand{\bea}{\begin{eqnarray}}
\newcommand{\eea}{\end{eqnarray}}

\shorttitle{Dynamical Friction in Cuspy Galaxies.}
\shortauthors{Arca-Sedda, Capuzzo-Dolcetta}

\begin{document}

\title{Dynamical friction in cuspy galaxies.}
\author{M. Arca-Sedda and R. Capuzzo-Dolcetta}
\affil{Department of Physics, Sapienza, Universit\`{a} di Roma, P.le A. Moro 5, I-00185, Roma, Italy}

\begin{abstract}
In this paper we treat the problem of the dynamical friction decay of a massive object moving in an elliptical galaxy with a cuspidal inner distribution of the mass density. 
We present results obtained by both self-consistent, direct summation, $N$-body simulations, as well as by a new semi-analytical treatment of dynamical friction valid in such cuspy central regions of galaxies. A comparison of these results indicates that the proposed semi-analytical approximation is the only reliable in cuspy galactic central regions, where the standard Chandrasekhar's local approximation fails, and, also, gives estimates of decay times that are correct at $1\%$ respect to those given by $N$-body simulations. 
The efficiency of dynamical friction in cuspy galaxies is found definitively higher than in core galaxies, especially on more radially elongated satellite orbits.
As another relevant result, we find a proportionality of the dynamical friction decay time to the $-0.67$ power of the satellite mass, $M$, shallower than the standardly adopted $M^{-1}$ dependence.

\end{abstract}


\section{Introduction}

Gravitational encounters between a massive body and a sea of light particles, such as a globular cluster moving in a galaxy, leads to a braking of the motion of the satellite widely known as dynamical friction.

Dynamical friction plays a crucial role in several astronomical contexts: from large scales, since it drives the motion of galaxies in galaxy clusters, to smaller scales, due to the consequences of this mechanism on the motion of BHs and star clusters in galaxies. 

The satellite mass and its orbit, togheter with the geometry of the system in which it moves, are relevant in the determination of the braking effect. As an example, the geometry of the galaxy plays a crucial role leading to significant different efficiency of this mechanism in spherical, axysimmetric and triaxial galaxies \citep{Cha43I,Bin77,Pes92}. Moreover, the presence of a cusp in the background matter distribution could affect it \citep{Merri,Vicari07}.

Actually, the existence of \lq cuspy\rq~density profiles of matter in galaxies has been argued in the last 20 years as a result of high resolution observations by the Hubble Space Telescope. 

Many galaxies exhibit, indeed, in the inner region, a luminosity profile steeply increasing toward their geometrical center, at least within the telescope resolution.

 In general, these luminosity profiles are well described by the so called S{\'e}rsic profiles \citep{Ser63}

\begin{equation}
\ln I(R)=\ln I(0) -k R^{1/n},
\label{sers}
\end{equation}
 
where $R$ is the projected radial coordinate and $n>0$, called the \lq S{\'e}rsic index\rq~, controls the steepness of the profile. Brighter galaxies have larger best fit values of $n$ ($n=4$ corresponds to the \cite{Dev} fit to giant ellipticals profiles); dwarf galaxies are characterized by smaller values of $n$. 

Defining $\Gamma$ as the logarithmic derivative of the luminosity profile:

\begin{equation}
\Gamma (R) =\frac{\rm{d}\ln I}{\rm{d}\ln R}, 
\end{equation}

the brightness profile slope of S{\'e}rsic's model is

\begin{equation}
\frac{\rm{d} I}{\rm{d} R}= \frac{I(R)}{R} \Gamma (R) = - I(R) \frac{k}{n}R^{(1-n)/n},
\end{equation}

so that $n>1$ correspond to a \lq true\rq~cuspidal central brightness profile
($dI/dR \rightarrow -\infty$ for $R\rightarrow 0$).

In the innermost ($3-10$ arcsecs) regions of early type galaxies, it has been shown recently that the luminosity profile is well approximated by the core-S{\'e}rsic profile \citep{Graham04, Dullo12}:
\begin{equation}
I(R)=I'\left[1+1\left(\frac{R_b}{R}\right)^\alpha\right]^{\gamma/\alpha}\exp\left[-b\left(\frac{R^\alpha+R_b^\alpha}{R_e^\alpha}\right)^{1/(\alpha n)}\right],
\end{equation}
where $I'$ is given by:
\begin{equation}
I'=I_b2^{-\gamma/\alpha}\exp[b(2^{1/\alpha}R_b/R_e)^{1/n}],,
\end{equation}
$I_b$ is the luminosity evaluated at the break radius $R_b$, $\gamma$ is the slope of the inner power-law region, $\alpha$ regulates the transition between the power-law and the S{\'e}rsic profile. Moreover, $R_e$ is the half-light radius and $b$ is a function
of the shape parameter $n$ and is defined such to ensure that $R_e$ actually encloses half of the total luminosity.

Of course, the real existence of cuspy (infinite density) innermost profiles for galaxies is just an extrapolation below the resolution limit of the behaviour of the observed distribution. 
On a theoretical point of view, numerical simulations of standard Cold Dark Matter (CDM) halo dynamics predict density profiles with $\rho \propto r^{-1}$ at small radii \citep{NFW96}; this prediction does not depend on particular cosmogonies or choice of initial conditions \citep{Hus99a,H99b} or on the specific form of the dark matter power spectrum \citep{Eke01}.
Adding a dissipative baryon component makes mass distributions even more concentrated \citep{Blu86,Dub94}. 
Anyway, there is not a general consensus about the real existence of a cusp in the dark matter distribution, because it could be an artefact of the finite resolution of the $N$-body simulations. 
While the CDM scenario is surely working on large scales, on smaller scales it meets problems, because observations seem to indicate that faint galaxies have cored profiles instead of real cuspy innermost densities (also the observed underabundance of dwarf satellites of large galaxies is a problem in the CDM scenario). 

For this, and others, reason we notice that some authors proposed, relatively recently, finite density profiles as the Einasto models \citep{eina65} as better suited to describe dark matter haloes (for a deep discussion about this matter see for example \cite{Me06}).

An important additional point is that many (if not all) galaxies host at their center a Compact Massive Object, identified with a Super Massive Black Hole (SMBH), in massive galaxies (well above $10^{11}$M$_\odot$), or a Nuclear Star Cluster (NSC), in lower mass galaxies (around or below $10^{9}$M$_\odot$).
There are quite a few cases of galaxies where an SMBH coexists with a NSC.

Such objects shape the density profile of the host in the innermost region, down to and below $\sim 10pc$. At present, it is still unclear whether NSCs have a cuspy density profile. Actually, while it is ascertained that some NSCs have a cored profile, as in the case of  the Milky Way NSC \citep{Do13}, it is not yet clear what is the innermost region of the majority of galactic nuclei because of resolution limits.

In this paper we study a possible solution of the problem of giving a reliable, quantitative, estimate of the dynamical friction effect on massive objects moving in a background of matter whose density profile is described by a cuspy central distribution.

This paper is organized as follows: in Sect. 2 we present the problems arising when applying the classic Chandasekhar formula for dynamical friction to the case of centrally diverging galactic density distributions and give a possible solution; Sect. 3 contains the results of the calibration of the semi-analytical approach of Sect. 2 by mean of $N$-body simulations; Sect. 4 presents the results obtained by our previously described approach regarding the actual modes of decay of massive objects in spherical, cuspy galaxies, also in  the presence of a massive central galactic black hole. Conclusions are drawn in Sect. 6. 

\section{Dynamical friction}
\label{sec2}
Dynamical friction (df) indicates the collective deceleration exerted on a massive body by the fluctuating force field where it moves. The existence of such effect was demonstrated by \cite{CVN42,CVN43} in their pioneering studies on this subject. 
Further on, \citet{Cha43I,Cha43b} developed a theory of dynamical friction which leads to a quantitative estimate of the braking in the simplified scheme of an infinite and homogeneous distribution of field stars.

In the astronomical context, the fluctuating force is given by the gravitational encounters between the test mass and the field objects, assumed as stochastic events. These encounters become significant over the mean field whenever they are close enough (strong encounters) or when the cumulative effect of weak encounters has grown sufficiently. The interest in the study of dynamical friction  in astronomy is on various sides. For instance, dynamical friction can be the way to accumulate matter in the inner regions of galaxies, so to explain, for instance, the formation of central Compact Massive Objects \citep{Dolc93,DoMioA,DoMioB}.
An observable consequence of this braking process is the evolution of the Globular Cluster System radial distribution in their hosting galaxies: the dynamical erosion would cause a flattening of the GCS radial ditribution around the center of the galaxy, as actually seen in many galaxies (see discussions in \cite{CaDV97},
 \cite{CDD01}, \cite{CDMB09}).

Moreover, dynamical friction determines the decay of Super Massive Black Holes (SMBHs) in remnants of merged galaxies \citep{Milos01} that leads to the formation of SMBH binaries. As a consequence of gravitational wave emission, the binary shrinks until the merging of the two components \citep{Schutz99}. In a non spherical merging, the final SMBH gains a kick that pulls the object out of its original position \citep{Bekenstein}, and the kicked BH could escape from the galaxy \citep{Campanelli07}. However, for small kick velocity, the recoiled object tends to decay again into the galactic centre because of dynamical friction \citep{Gual08,Vicari07}. Also the commonly observed presence of a giant elliptical galaxy at center of galaxy clusters is attributed to this dynamical deceleration whose action is stronger on more massive galaxies.

The effect of dynamical friction in a galaxy depends on both the orbit of the massive test object and on the local phase space density along this orbital path. 
Regarding the overall matter distribution of the host galaxy, the lack of symmetry in the potential favours dynamical braking because of the loss of angular momentum conservation and the consequent closer approach of the massive objects to the central denser region of the galaxy \citep{Pes92}. Moreover, it is known that the central galactic regions are those of highest phase space density (as measured by the proxy $\rho/\sigma^3$) so to make low eccentricity orbits as the ones suffering most of the deceleration. Consequently, central regions of cuspy galaxies with a triaxial shape over the large spatial scale are candidates to be sites of strongly enhanced dynamical decay, so as to convince us of the importance of its correct evaluation.

\subsection{Is dynamical friction deceleration diverging in the central density cusp?} 

\begin{figure}
\centering
\includegraphics[width=10cm]{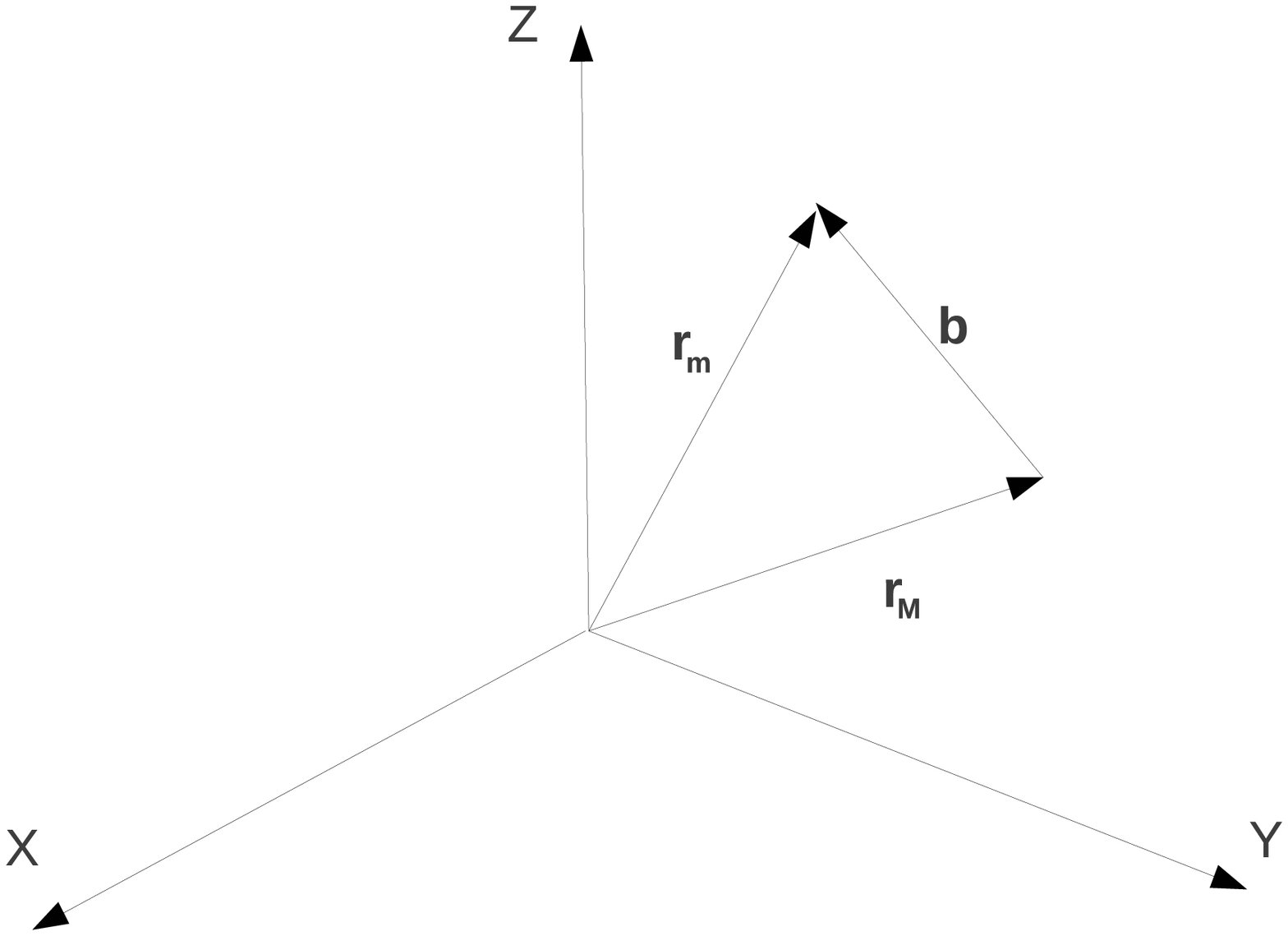}
\caption{The symbol $\textbf{r}_M$ indicates the position vector of the test particle of mass $M$, while $\textbf{r}_m$ is the position of the field particle of mass $m$; $\textbf{b}$ indicates the impact vector pointing to the field particle.}
\label{system}
\end{figure}

Letting $M$ and $m$ the mass of the test particle and of the generic field star, respectively, and identifying with $\mathbf{v}_M$ and $\mathbf{v}_m$ their velocities, given also the impact parameter vector $\bf{b}$ (see Fig.\ref{system}), the 2-body hyperbolic interaction between the test mass and the field star induces the velocity variation for the test mass:
\begin{equation}
	\Delta \mathbf{v}_M = -\left(\frac{m}{m+M}\right) 2V \left[ 1+\frac{b^2V^4}{\mathrm{G}^2(m+M)^2} \right]^{-1} \frac{\mathbf{V}}{V},
\label{hypdefl}
\end{equation}

where $\mathrm{G}$ is the Newton's gravitational constant and $\mathbf{V}=\mathbf{v}_M-\mathbf{v}_m$ is the 2-body relative velocity.

The effective time duration of such a 2-body interaction is the {\it fly-by} time, assumed to be $\Delta t \sim 2b/V$, so that the mean deceleration due to the single encounter in the direction of the initial motion is well approximated by $\Delta \textbf{v}_M / \Delta t$. Consequently, the global deceleration effect is simply given by an integral over the whole distribution of scatterers:

\begin{equation}
\left(\frac{\rm{d}\mathbf{v}_{\it M}}{\rm{d} t}\right)_{\rm{df}}= \int \frac{\Delta \mathbf{v}_M}{\Delta t}\mathrm{d}N,
\label{dfdefin}
\end{equation}

where $\mathrm{d}N$ is the (infinitesimal) number of field stars in the elementary space volume centered in $\mathbf{r}_m = \mathbf{r}_M + \bf b$ having velocities in the (infinitesimal) velocity volume centered in $\bf{v}_m$. Once that the field stars' steady state distribution function 
is known as $f(\mathbf{r}_m,\mathbf{v}_m)$,  $\mathrm{d}N$ is written as:

\begin{equation}
\mathrm{d}N = f(\mathbf{r}_m,\mathbf{v}_m)\mathrm{d}^3 \mathbf{v}_m \mathrm{d}^3 
\mathbf {r}_m.
\end{equation}

As a consequence, we can express the mean cumulative deceleration in Eq. \ref{dfdefin} as: 

\begin{equation}
\left(
\frac{\mathrm{d}\mathbf{v}_{\it M}}{\mathrm{d}t}
\right)_{\mathrm{df}} =
-\frac{m}{m+M}
\int \!\! \int 
f(\mathbf{r}_M+\mathbf{b},\mathbf{v}_m)\frac{V}{b}
\frac{\mathbf V}{1+ b^2V^4G^{-2}(M+m)^{-2}}
\mathrm{d}^3\mathbf{v}_m \mathrm{d}^3\mathbf{b},
\label{dfcorrect1}
\end{equation}
where $\mathbf{r}_m=\mathbf{r}_M+ \mathbf{b}$ and the integral is over the whole range of values of $\bf b$ and $\bf v_m$ allowed by self-consistency.

The integral in Eq.\ref{dfcorrect1} is, in general, quite complicated.

It can be performed in the hypothesis of a DF separated in the space and velocity coordinates, $f(\mathbf{r}_m,\mathbf{v}_m)\equiv g(\mathbf{r}_m)G(\mathbf{v}_m)$, and the additional \lq local\rq~approximation, which means that in the evaluation of the integral in Eq.\ref{dfcorrect1} the substitution of $g(\mathbf{r}_m)$ with $g(\mathbf{r}_M)$ is done, that corresponds to weighting  encounters at any generic distance $\mathbf{r}_m$ from the test object with the local density (i.e. where the satellite is, $\mathbf{r}=\mathbf{r}_M$). The local approximation allows an integration over the impact parameters which leads to the function

\begin{equation}
\frac{1}{Q^2}\ln\left(1+Q^2b_{max} \right)
\label{qfunc}
\end{equation}

where $Q^2=Q^2(V)=V^4/[G(m+M)]^2$ and the cut at $b=b_{max}$ is needed to avoid the logarithmic divergence. 
Letting $\Lambda\equiv 1+Q^2b_{max}$, the further assumption of isotropy in the velocity dependent function leads to the simple expression:

\begin{equation}
\left(\frac{\rm{d}\mathbf{v}{\it _M}}{\rm{d}t}\right)_{loc}=-4\pi^2 G^2(m+M) \ln\Lambda \rho(r_M,v_m\leq v_M)\displaystyle \frac{\mathbf{v}_M}{v_M^3},
\label{local}
\end{equation}

where $\rho(r_M,v_m\leq v_M)$ is the, local, mass density of field stars slower than the test particle.
At this last regard, it may be worth noted that often the logarithmic function of field star velocities is taken out of the integral also out of the previous assumptions, when it can correctly take out. This happens, for instance when, in spherical simmetry, a distribution function in the form $f(\mathbf{r}_m,\mathbf{v}_m)\equiv f(r_m,v_m)=f(E)$, where $E=v_m^2/2+\Phi(r_m)$ is the field stars mechanical energy per unit mass, is assumed. This simplification, valid whenever the function in Eq.\ref{qfunc}  does not vary significantly over the allowed velocities, implies 

\begin{equation}
\rho(r_M,v_m\leq v_M)= 4\pi \int_0^{v_M} v_m^2 f[v_m^2/2+\phi(r_M)] \mathrm{d}v_m.
\end{equation}

Several authors (e.g. \cite{Trem76}) suggested that allowing a variation of the 
Coulomb logarithm, $\ln \Lambda$, may be important for a good determination of its orbital evolution. \cite{JP05} derived an expression for $\ln \Lambda$ allowing the variation of the maximum impact parameter, $b_{max}$, and the $a_{90°}$ parameter (the typical impact parameter for a $90°$ degrees deflection in a two body encounter). The effect of this variations on the orbits of massive body traveling in cuspy galaxies are deeply discuss in \cite{Spurz}.

Now, whenever the test particle is significantly off center with respect to the stellar system (star cluster, galaxy, etc.) where it moves, the local expression (Eq. \ref{local}) gives an acceptable approximation; on the contrary, it loses its validity in the neighbourhood of the host system center. In this case the local approximation is clearly an overestimate of the actual dynamical friction, because it corresponds to weighing the contribution of the gravitational encounters at any distance from the test particle not with the, correct, density of target stars at that distance but, rather, with the density of targets evaluated at the location of the test particle
itself, that is maximum at the origin of any self-gravitating system.
This overestimate is a particularly serious problem when dealing with cuspy galaxies, where the spatial density of stars {\it diverges} at the galactic center.
This divergence may be partially cured by introducing an artificial spatial cut-off in  the density distribution but this, of course, implies a relevant dependence of dynamical friction on the choice of this radial cut-off.

We can illustrate better all this with the example of a distribution function as obtained using
a $\gamma$ model \citep{Deh93} around the central spatial cusp of a spherical galaxy, where the stellar density may indeed be represented as
$\rho(r) \propto r^{-\gamma}$.  
As it is easily seen (see Appendix A), when $\gamma=1$ the following expression for 
$\rho(r,v_m\leq v_M)$ in the high binding energy regime, i.e. around the galactic center, is obtained:

\begin{equation}
\rho(r,v_m\leq v_M)= \frac{4\pi}{3}A\frac{v_M^3}{(r/a)\left[
(v_M^2/2)/(GM/a)+r/a\right]^{3/2}}
\label{dfhbeg=1}
\end{equation}

where $A$ is the multiplicative constant in the expression of the distribution function (see Appendix A). 
The resulting local approximation (Eq. \ref{local}) for the dyn deceleration yields

\begin{equation}
\left(\frac{\mathrm{d}\mathbf{v}_M}{\mathrm{d}t}\right)_{loc}
=-\frac{16\pi^3}{3} AG^2(m+M) \ln \Lambda \frac{\mathbf{v_M}}{(r/a)\left[
(v_M^2/2)/(GM/a)+r/a\right]^{3/2}}.
\label{dfcoeflg=1}
\end{equation}

If, in the denominator of Eq. \ref{dfcoeflg=1}, $r/a$ and $(v_M^2/2)/(GM/a)$ go (contemporarily) to zero with same order of infinitesimal, the local dyn deceleration  diverges as $(r/a)^{-2}$ (or, equivalently, $[v_M^2/(GM/a)]^{-2}$).
This divergence is due to the local approximation, while the correct (Eq. \ref{dfcorrect1}) expression for the deceleration {\it does not} diverge; on the contrary, deceleration goes to zero for particles of very high binding energy (see Appendix A). Therefore the local approximation formula cannot be used to get astrophysically significant results when treating 
the motion of massive objects passing through (or close to) the center of a cuspy galaxy.

\subsection{A possible solution of the divergence problem}
\label{divergence}
In Appendix A we show that the fully isotropic distribution functions of the Dehnen's gamma model lead to a deceleration which is finite around the galactic central density cusps, while its local approximation is not.
This convinces us of the need to use, instead of the wrong local approximation, the complete (Eq. \ref{dfcorrect1}) expression for the dynamical friction process. 

Unfortunately, the integral in Eq. \ref{dfcorrect1} is of overwhelming complexity, unless some simplifications are adopted. An intuitive, immediate, approximation comes from letting both ${\bf r}_M =0$ and spatial isotropy (i.e. spherical symmetry) for the distribution function ($f({\bf r}_m, {\bf v}_m)=f(r_m,{\bf v}_m)$) to get the simpler expression for the deceleration:

\begin{equation} 
\left(\frac{\mathrm{d}\mathbf{v}_{\it M}}{\mathrm{d}t} \right)_{\mathrm{cen}} = -{{4\pi m}\over {m+M}}
\int_{b_{min}}^{b_{max}} \! \! \! \int 
\! f(b,\mathbf{v}_m)
\frac{V}{1+b^2V^4G^{-2}(m+M)^{-2}} 
\mathbf{V} \mathrm{d}^3 \mathbf{v}_m b \mathrm{d}b,  
\label{dfcen1}
\end{equation}

where $b_{min}$ and $b_{max}$ are, respectively, the minimum and maximum impact parameters allowed. The lower minimum cannot be zero, because this would correspond to a front collision, i.e. to a radial relative motion which does not fulfil the basic condition of positive mechanical energy for the idealized 2-body encounter. On the other side, the upper limit, $b_{max}$, is, usually, chosen large enough to guarantee that the stellar density at distance $b_{max}$ from the center is much smaller than in the neighbourhood of the test object.\\
For a huge set of distribution functions, the vector integral in Eq. \ref{dfcen1} is both convergent (see Appendix A) and suited to a proper numerical integration. 
The integration over field stars' velocities in Eq. \ref{dfcen1} has been done over the 
all interval allowed, i.e. limited to the central escape velocity.

Of course, the dynamical friction evaluated this way gives a good result along the motion of the test mass in the neighbourhood of the galactic center but cannot be used on a larger spatial scale. Consequently, our choice was that of an interpolation between the `central' dyn evaluation and the `local' approximation, by mean of a proper interpolation formula of the type:

\begin{equation}
\left(\frac{\mathrm{d}\mathbf{v}_M}{\mathrm{d}t}\right)_{\mathrm{df}}=p(r)\left(\frac{\mathrm{d}\mathbf{v}_M}{\mathrm{d}t}\right)_{cen}+\left[1-p(r)\right] \left(\frac{\mathrm{d}\mathbf{v}_M}{\mathrm{d}t}\right)_{loc},
\label{interp}
\end{equation}

where the interpolation function, $0\leq p(r)\leq 1$, is assumed monotonically decreasing from $p(0)=1$ outward. Within these constraints, the interpolation function is a priori arbitrary; the only way to tune it is through a careful comparison with $N$-body simulations of the decay of massive objects under different initial conditions. 
Thanks to this comparison, we found that a good interpolation expression is $p(r)=e^{-r/r_{cr}}$ where $r_{cr}$ is the size of the region of dominance, in the contribution to the dynamical friction, of the  central cusp. 
The actual $r_{cr}$ values are determined in Sect. \ref{calibration}.
It is relevant noting that although the exponential choice is not unique, the, simpler, linear interpolation can be excluded because our results shows that a linear function weights too much the central contribution, giving an unrealistically high deceleration. 

The computation of the scattering integral in Eq. \ref{dfcen1} 
presents numerical difficulties due to the singularity in the integrand. These difficulties can be overcome by using a proper integration algorithm; in particular, we used DECUHR, an algorithm which combines an adaptive subdivision strategy with extrapolation \citep{genz}. 

\begin{figure}
\centering
\includegraphics[width=8cm]{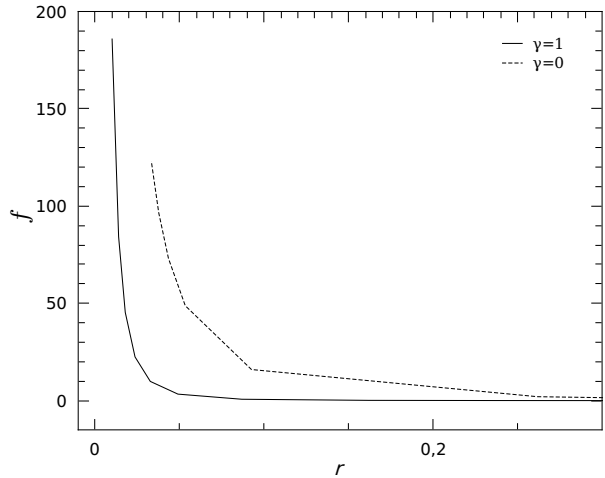}
\caption{The ratio, $f$, of the dynamical friction acceleration evaluated with the local approximation formula in Eq. \ref{local} to the central given by Eq. \ref{dfcen1}, in the cases $\gamma =0$ (dashed line) and $\gamma=1$ (solid line).}
\label{acc_Dv}
\end{figure}

In Fig. \ref{acc_Dv} it is evident the departure of the local friction evaluated via Eq \ref{local} respect to the central estimate given by Eq. \ref{dfcen1}.

We can note that, in the regime of very low or very high speed for the test particle and/or when its mass, $M$, is small, the above integration algorithm requires an exceedingly large number of iterations to reach convergence. In such  cases, to speed up computations we looked for an appropriate approximation formula.

We actually found that the linear dependence of dyn on $v_M$ is recovered, while at high velocities the dependence is a power law with a spectral index, $\alpha$, that depends both on $\gamma$ and on $b_{min}$

\begin{equation*}
\alpha=\begin{cases} 
2(\gamma-1) & \text{if $b_{min}=0$,}
\\
-2 & \text{if $b_{min}>0$,}
\end{cases}
\end{equation*}

in the range $0\leq \gamma \leq 2$.

\section{Calibration by mean of $N$-body simulations}
\label{calibration}

A fully self consistent study of the dynamical friction caused by environment stars on the motion of a (massive) object 
of mass $M$ requires the numerical integration of an $N$-body problem where $N_f$ particles sample the galactic field and $N_M$ particles represent the massive star system ($N_f+N_M=N$). In principle, to have results of high reliability in the astrophysical context, high resolution simulations are needed, which require both a large value of $N_f$ and of $N_M$. 

This may be unfeasible when aiming to a statistically complete set of simulations over a huge set of initial conditions. On the other hand, an analytical, or semi-analytical approach, although much more suited to an extensive analysis suffers of its intrinsic, more or less severe approximations. The natural way to treat in a simplified scheme the topic of dynamical friction of massive objects in a stellar bacgkground is that of the integration of the equations of motion of the single, massive, object in a given external potential $\Phi(\textbf{r})$ with the inclusion of the drag term given by Eq. \ref{interp} ( dragged one-body problem).

Taking all this into account, a good choice can be that
of a proper calibration of the free parameters in the dragged one-body problem by mean of a set of reliable, high precision $N$-body simulations.

\subsection{The dragged one-body problem}
In the dragged one-body problem, the equations of motion to solve are written as:

\begin{equation}
\ddot{\mathbf{r}}_{M}=\nabla\Phi(\mathbf{r}_M) + \left(\frac{\mathrm{d}\mathbf{v}_{\it M}}{\mathrm{d}t}\right)_{\mathrm{df}},
\label{onebody}
\end{equation}

with the proper initial conditions.
To solve this set of differential equations, we use a high precision $6^{th}$($7^{th})$ order Runge-Kutta-Nystr\"om method with variable time step \citep{Feh72}. 
The time step size, $\Delta t$, was varied according to  
\begin{equation}
    \Delta t = \eta\min\left(
    \frac{|\mathbf{r}_M|}{|\dot{\mathbf{r}}_M|},
    \frac{|\dot{\mathbf{r}}_M|}{|\ddot{\mathbf{r}}_{M}|}\right), 
\nonumber 
\end{equation}
 
that, with the choice of $\eta=0.01$, allows both a fast integration and an energy and angular momentum conservation at a fractionary $10^{-11}$ level (per time step).

In this paper we choose as units of mass and length the galactic mass and scale length
of its density distribution, denoted by $M_G$ and $a$. The further choice of
setting the gravitational constant $\rm{G}=1$ leads to 
 
\begin{equation}
T=\frac{a^{3/2}}{\sqrt{\rm{G}M_G}}
\label{PHt}
\end{equation}

as unit of time.

Once that the expression for the interpolation function, $p(r)=e^{-r/r_{cr}}$ cited in Sect. \ref{divergence}, is given, the free parameters in the semi-analytical evaluation of dynamical friction are the scale length $r_{cr}$ in $p(r)$, and the values for $b_{min}$ and $b_{max}$ in the local (Eq.\ref{local}) and central (Eq.\ref{dfcen1}) expressions of the deceleration. 

We made several simulations using both constant and variable $b_{max}$, to conclude that the advantages in accuracy given by a somewhat arbitrary variation in $b_{max}$ are not such to overcome the simplicity of the choice of $b_{max}$ set at the constant value $R$, the assumed radius of the spherical galaxy. 
On the other side, due to its undoubted relevance in a cuspy galaxy, we let $b_{min}$ to vary. Also the length scale $r_{cr}$, which determines the size of the region of dominance of the central to the local friction term, is allowed to vary.

An unambiguous way to select their optimal values is through a comparison of results got via the integration of Eq. \ref{onebody} at varying the pair ($r_{cr}$,$b_{min}$) and the, supposedly `exact', results coming from the integration of motion of a single, point-like, massive object of mass $M$ interacting with $N$ bodies of mass $m$ representing the galactic field. At this scope we used our direct summation, high precision, $6th$ order Hermite's integrator with individual block time steps called HiGPUs \citep{CDSP13a}. HiGPUs runs on composite platforms where the host governs the activity of Graphic Processing Units (GPUs) as computing accelerators. The code exploits all the potential of such architectures, since it uses at the same time Message Passing (MPI), Open Multiprocessing (Open MP) on the host CPUs and Compute Unified Device Architecture (CUDA) or Open Computing Language (OpenCL) on the GPUs \citep{CDSP13b}.
HiGPUs has been extensively checked also in its accuracy; anyway, for the purposes of this paper (once that the optimal number of particles has been set) we performed several simulations to check its accuracy. In particular, we verified that over the time lentghs of relevamce for our aims, the code conserves the total energy, linear and angular momentum of the system with a relative error down to $10^{-8}$ (for energy) and down to $10^{-10}$ for momentum. Moreover, we checked that the simulated systems do not expand or contract significantly during their evolution, as a guaranty of both  correct choice of initial conditions and quality of time integration. The system stability has been verified also looking at the lagrangian radii and density profiles, which remain substantially constant during the whole orbital evolution of the satellite, a part from local wake effects induced by the satellite motion.

\subsection{Sampling effects}
\begin{figure}
\centering
\includegraphics[width=8cm]{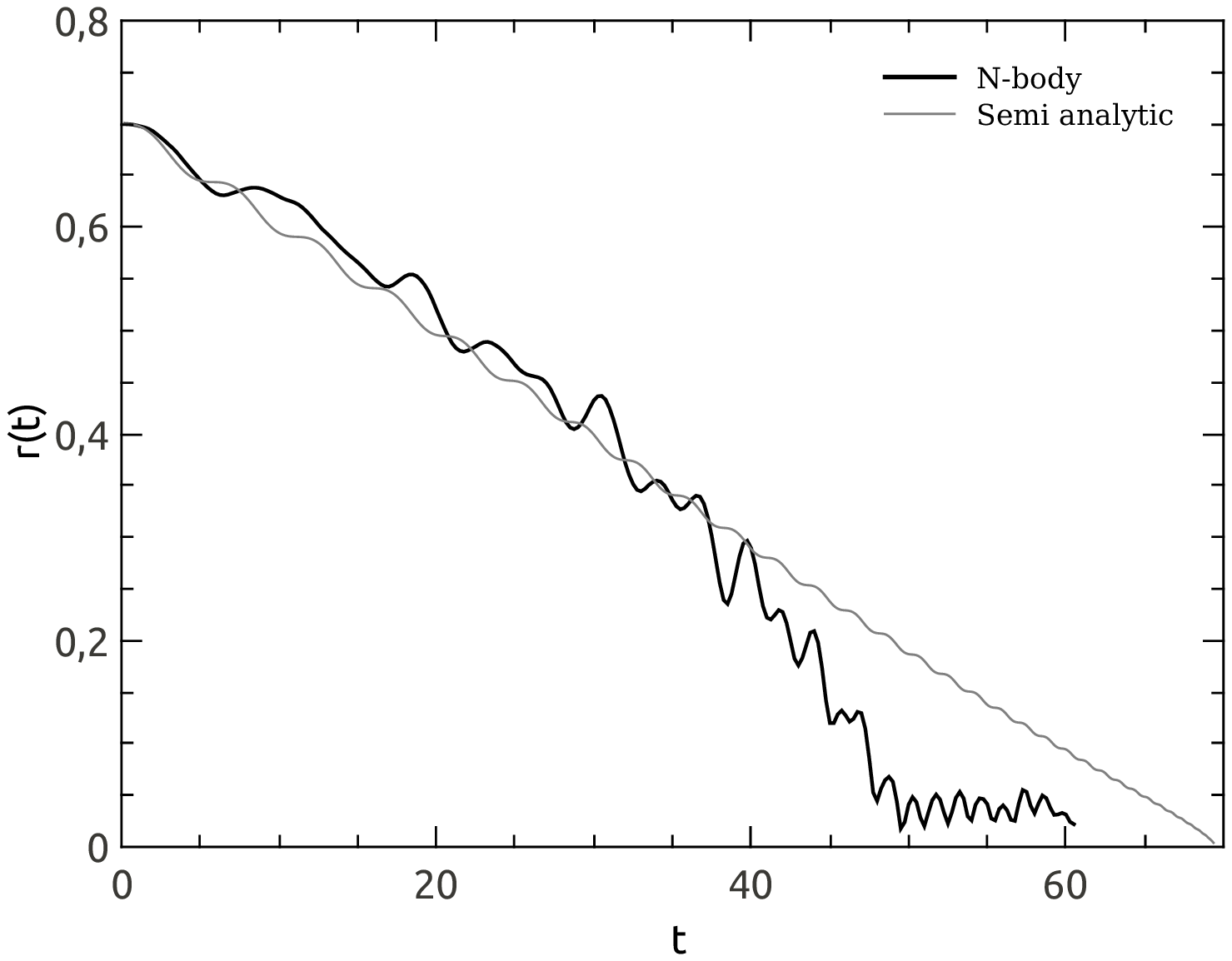}
\caption{Time evolution of the galactocentric distance of an $M=10^{-3}$ test mass on initially circular orbit in the $\gamma=1$ model.} 
\label{tr3}
\end{figure}

In order to make an optimal selection of the two free parameters needed to set the drag term in the one-body scheme, we perform an adequate set of direct $N$-body integrations, as explained above. To calibrate these parameters it is, of course, important to be sure of the reliability of such $N$-body simulations. The main problem, at this regard, is the sampling. Actually the $N$-body sampling acts on both small (\lq granularity\rq~) and large  (deviation from spherical symmetry) scale. This makes initially circular orbits evolve into precessing ellipses of moderate eccentricity (see Fig. \ref{tr3}). This is one of the, unavoidable, causes of departure of the decay times in the $N$ case from the semi-analytical case. 
To reduce spurious sampling effects we tried to determine an acceptable threshold value of $N$ above which fluctuations are kept small enough. To do this, we followed the orbital evolution of a particle of the same mass of the generic particle of the $N$-body representation of the galaxy, starting from initial conditions corresponding to the extreme (in eccentricity) cases of circular and radial orbits.

\begin{figure}
\centering
\includegraphics[width=8cm]{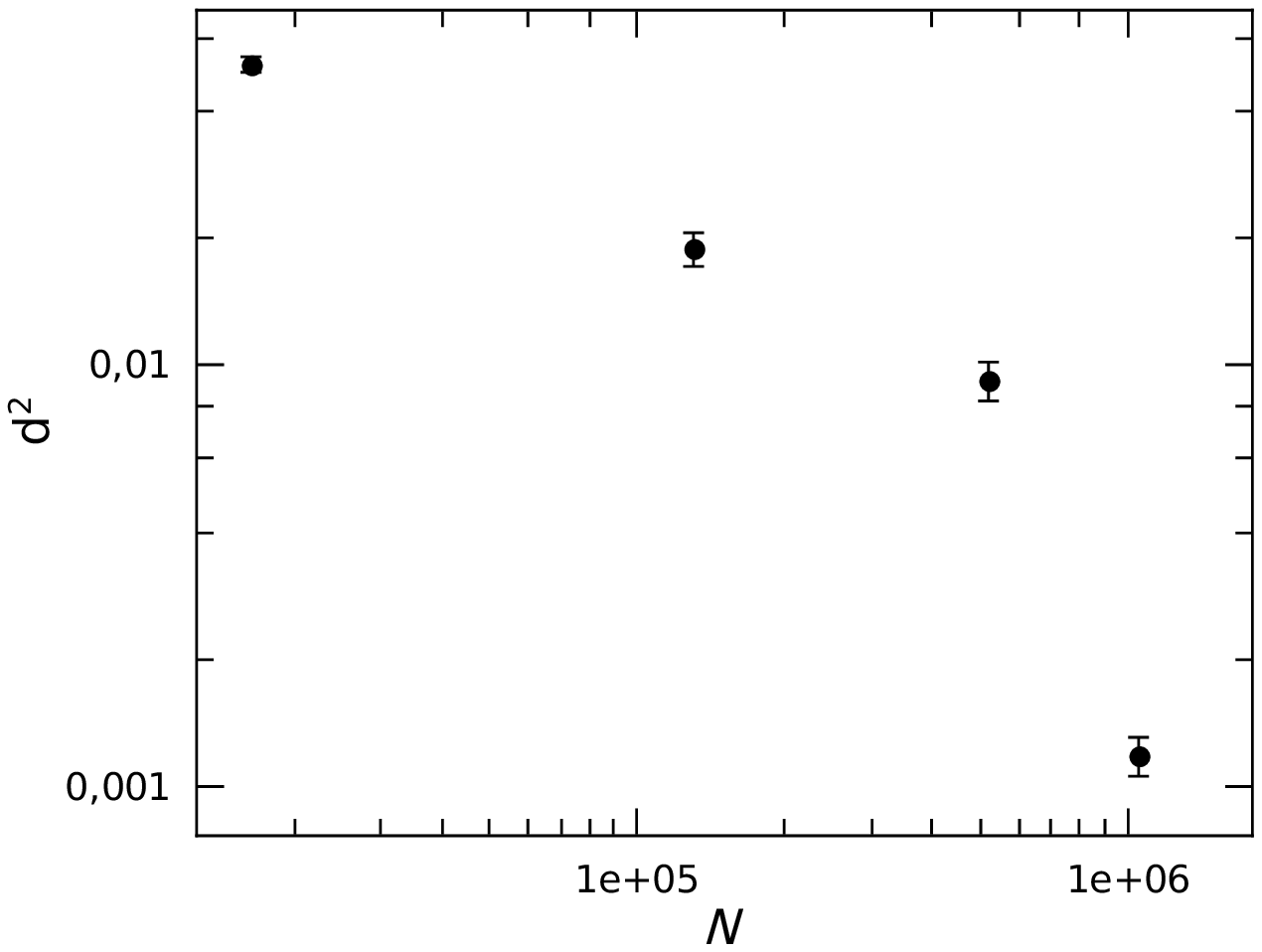}
\caption{The squared, fractional departure of the distance to the center of a test particle of same mass of the field particles along its motion as integrated in an $N$-body sampled $\gamma=1$ model respect
to the ideal radial orbit.}
\label{devrad}
\end{figure}

\begin{figure}
\centering
\includegraphics[width=8cm]{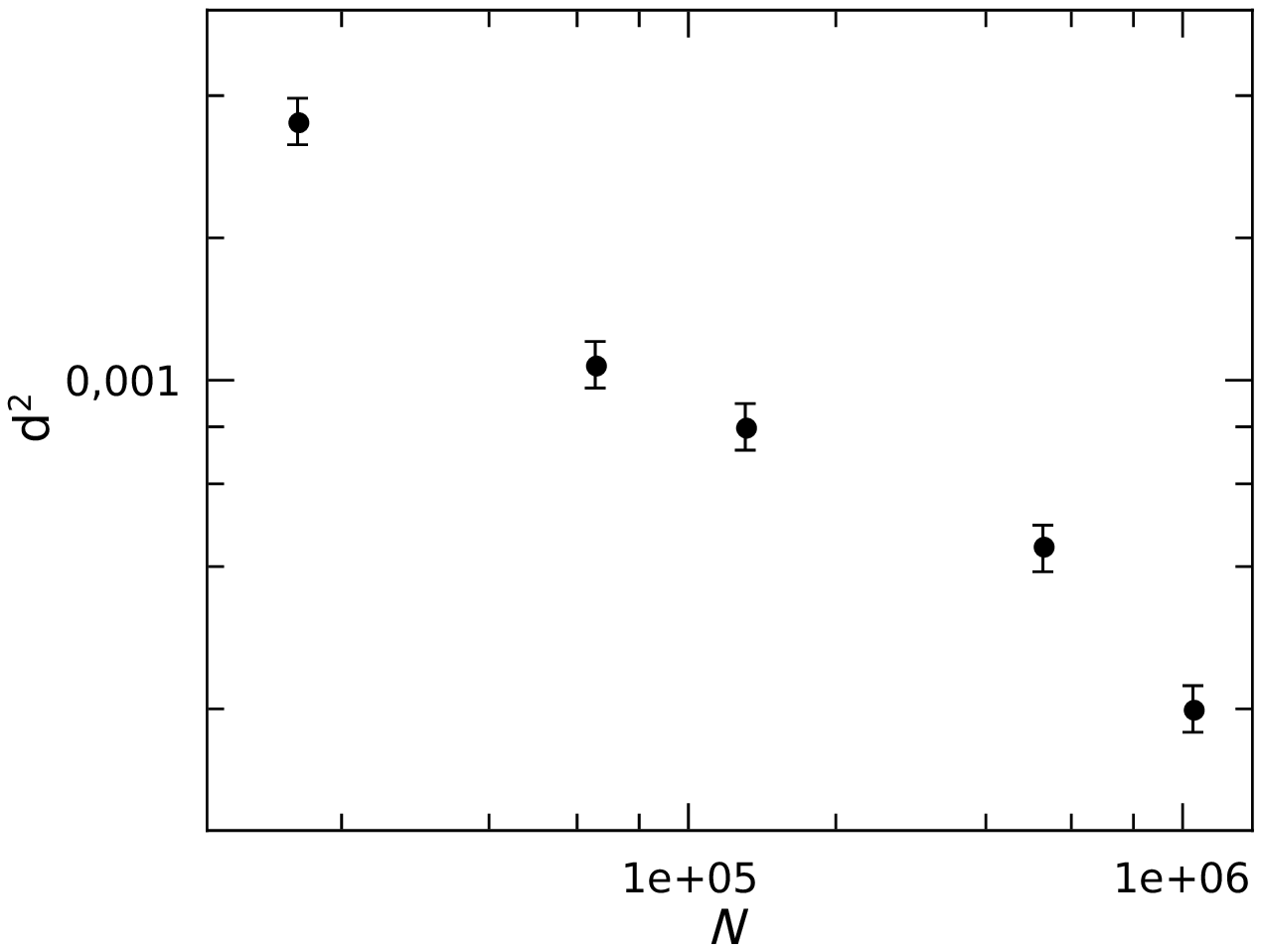}
\caption{The squared, fractional departure of the distance to the center of a test particle of same mass of the field particles along its motion as integrated in an $N$-body sampled $\gamma=1$ model respect
to the ideal circular orbit.}
\label{devcirc}
\end{figure}

As it can be seen from Figs. \ref{devrad} and \ref{devcirc}, in both of these extreme cases the quadratic deviation of the actual trajectory computed in a finite $N$-body representation of a Dehnen's $\gamma=1$ galactic density law 
respect to the ideal (infinite $N$) circular and radial trajectories decreases significantly when $N$ is in the range $10^5 <N< 10^6$.
Actually, the reduction of fluctuations passing from $N=131,072$ to $N=524,288$ suggested us to choose this latter value as a good compromise in giving an acceptable smoothness at a reasonable computational cost.

\subsection{Determination of the free parameters in the one-body scheme}

Once determined the threshold in $N$ over which an acceptable fit between the $N$-body test object integration and that obtained by the solution of the single body motion in the external smooth galactic field, the following step was that of getting reliable pairs of values $(r_{cr},b_{min})$ in dependence on $\gamma$. We proceeded this way:
 i) perform $N$-body integrations of the motion of a massive point mass, starting from an initial distance $r_{cr}$ from the galactic center with the local circular velocity; ii) perform a similar time evolution in the simplified one-body scheme of Eq. \ref{onebody} where the dynamical friction dissipation term is given in the standard local approximation form; iii) reduce $r_{cr}$ until the difference between the orbit self-consistently evaluated in i) and that obtained as explained in ii) changes significantly; iv) take this latter value of $r_{cr}$ as optimal value for the $p(r)$  function in the interpolation formula (Eq.\ref{interp}).
To do this we set $M=10^{-3}$ as value of the test particle mass.

\begin{figure}
\centering
\includegraphics[width=8cm]{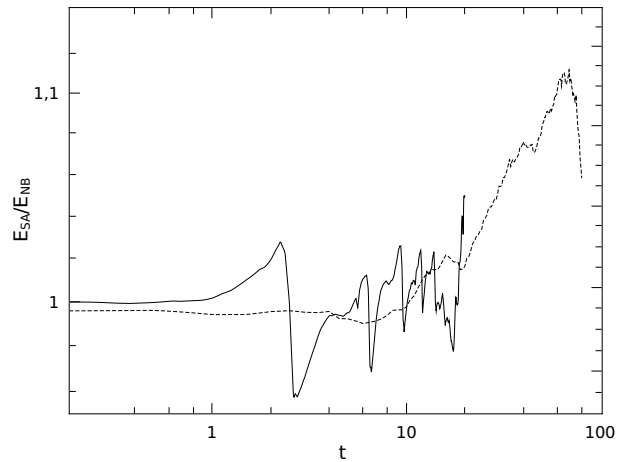}
\caption{Time evolution of the ratio between the test particle energy evaluated in the one-body, semi-analytical case and that computed in the $N$-body sampled galaxy for a radial (solid line) 
and a circular orbit (dotted line).}
\label{enratio}
\end{figure}

An idea of the quality of this fitting procedure to determine the pair ($r_{cr}$,$b_{min}$) 
in getting the results of interest here is given by Fig. \ref{enratio}. It reports the ratio between the test particle orbital energy evaluated in the one-body approximation with the dynamical friction term written in the complete (Eq. \ref{interp}) form to that computed in the full $N$-body simulation. As it is seen, the variations are within $4\%$ over 20 time units in the radial case and within $2\%$ in the circular case over the same time interval. In the circular case we extended the comparison up to 80 time units, finding a relative maximum of the fractional difference of about 12 $\%$.

We found that the greater the  $\gamma$ the smaller the $r_{cr}$, as expected. Actually, higher values of $\gamma$ represent steeper profiles toward the center, with a large part of the total mass enclosed within a relatively small radius. 
On the other side, less intuitive is the result of $r_{cr}$ as very similar to the radius enclosing $10\%$ of the mass of the system. A simple inversion of the mass-radius profile for Dehnen's models gives:

\begin{equation}
r(x_M)=\frac{x_M^{1/(3-\gamma)}}{1-x_M^{1/(3-\gamma)}}
\end{equation}

with $x_M=M(r)/M_G$. The value of $r(0.1)$ is found (see Tab.\ref{rcr}) in good agreement with those of $r_{cr}$ obtained in the way indicated above.

\begin{figure}
\centering
\includegraphics[width=8cm]{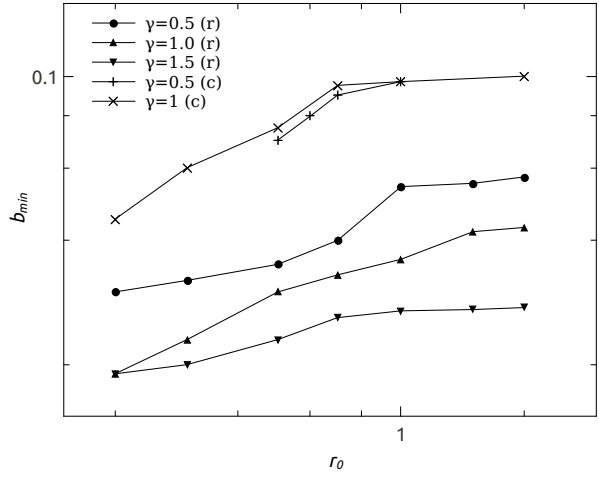}
\caption{Minimum impact parameter as a function of the initial galactocentric distance, for initially radial (r) and circular (c) orbits at various values of $\gamma$.}
\label{bmin}
\end{figure}

Once that the $r_{cr}$ values are obtained for different $\gamma$, to get the best minimum impact parameter $b_{min}$ we vary it in a set of one-body integrations covering circular and radial cases to find those best fitting results of direct $N$-body computations. In Fig. \ref{bmin} we show the  $b_{min}$ selected this way, as a function of $\gamma$.

\section{Results}

The main scope of this paper was to obtain reliable estimates of the role of dynamical friction in cuspy galaxies, as explained before.

This aim has been reached by means of both direct numerical integrations of the motion of a massive test particle in an $N$-body representation of the host cuspy galaxies and of the simpler, and much faster, one-body representation given by Eq. \ref{onebody} together with Eqs. \ref{local}, \ref{dfcen1}, and \ref{interp}.

In Tables \ref{RS1}, \ref{RS2}, and \ref{RS3} the fundamental data of the whole set of $N$-body simulations performed are given.

\begin{figure}
\centering
\includegraphics[width=8cm]{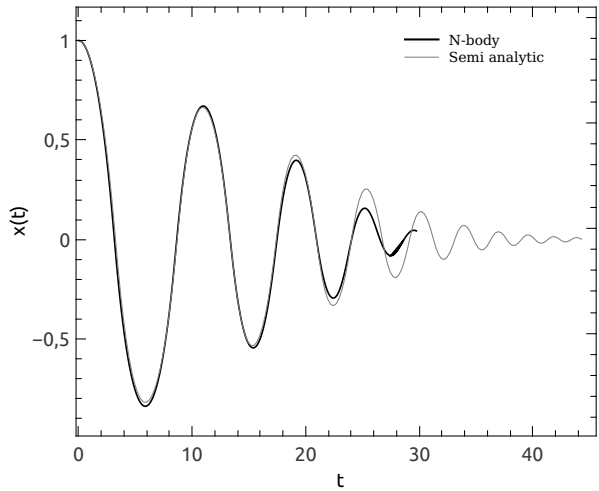}
\caption{Damped oscillations along the $x$ axis for the test object with mass $M=10^{-3}$  in the $\gamma=1$ model. The darker line refers to the $N$-body simulation, while the grey line to the semi-analytical.} 
\label{tr1}
\end{figure}

\begin{figure}
\centering
\includegraphics[width=8cm]{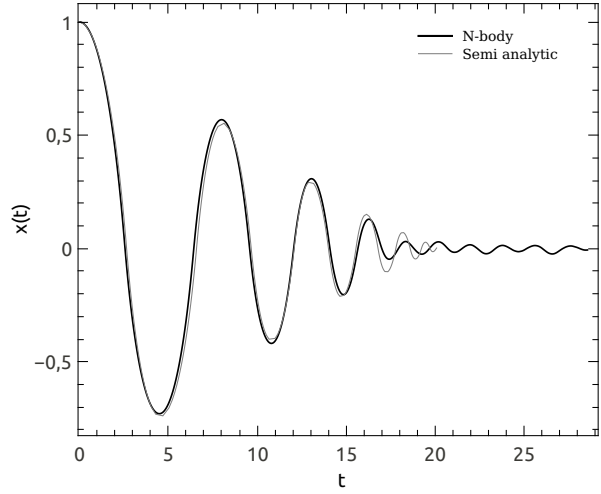}
\caption{As in Fig.\ref{tr1}, but for the model with $\gamma=1/2$.} 
\label{tr2}
\end{figure}

\begin{figure}
\centering
\includegraphics[width=8cm]{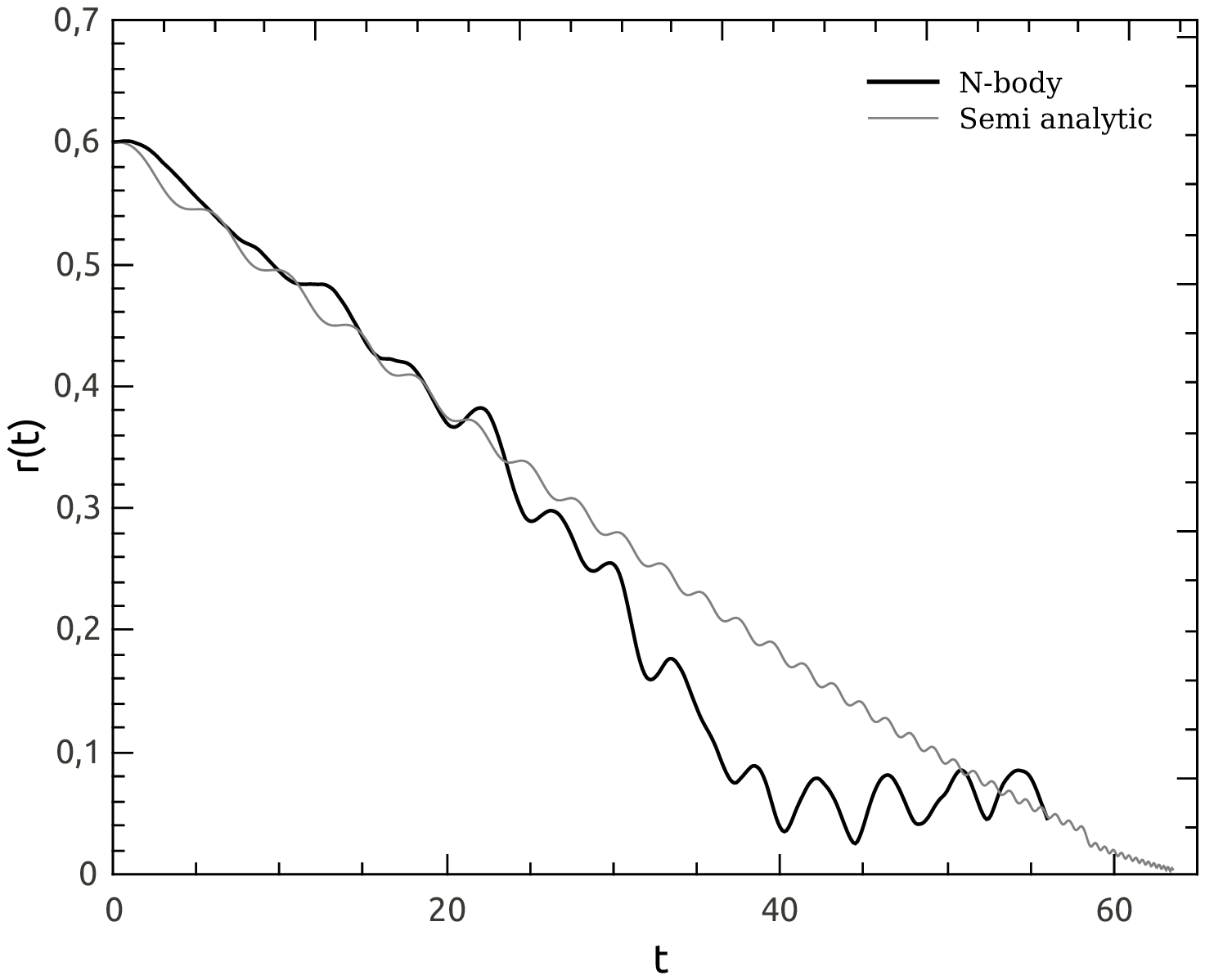}
\caption{Time evolution of the galactocentric distance of an $M=10^{-3}$ test mass on initially circular orbit in the $\gamma=1/2$ model.} 
\label{tr4}
\end{figure}

\begin{figure}
\centering
\includegraphics[width=8cm]{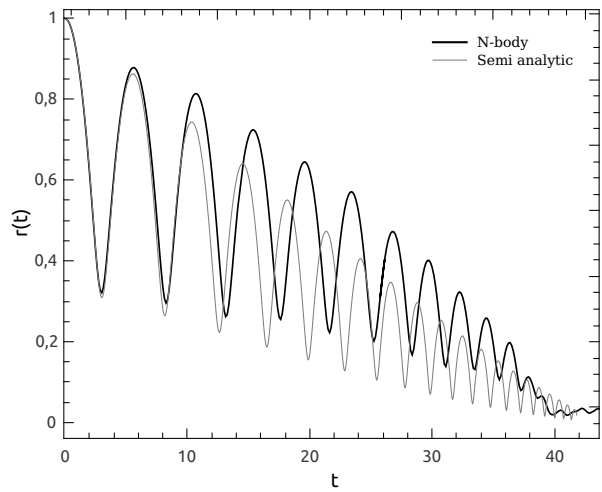}
\caption{Same as in Fig.\ref{tr4}, but for an eccentric orbit with $e=0.5$.} 
\label{tr5}
\end{figure}

Using these $N$-body simulations as reference, the quality of the one-body treatment is given in Figs. \ref{tr1}, \ref{tr2}, \ref{tr4}, and \ref{tr5}, 
where the time evolution of the test mass galactocentric distance is reported. 

\begin{figure}
\centering
\includegraphics[width=8cm]{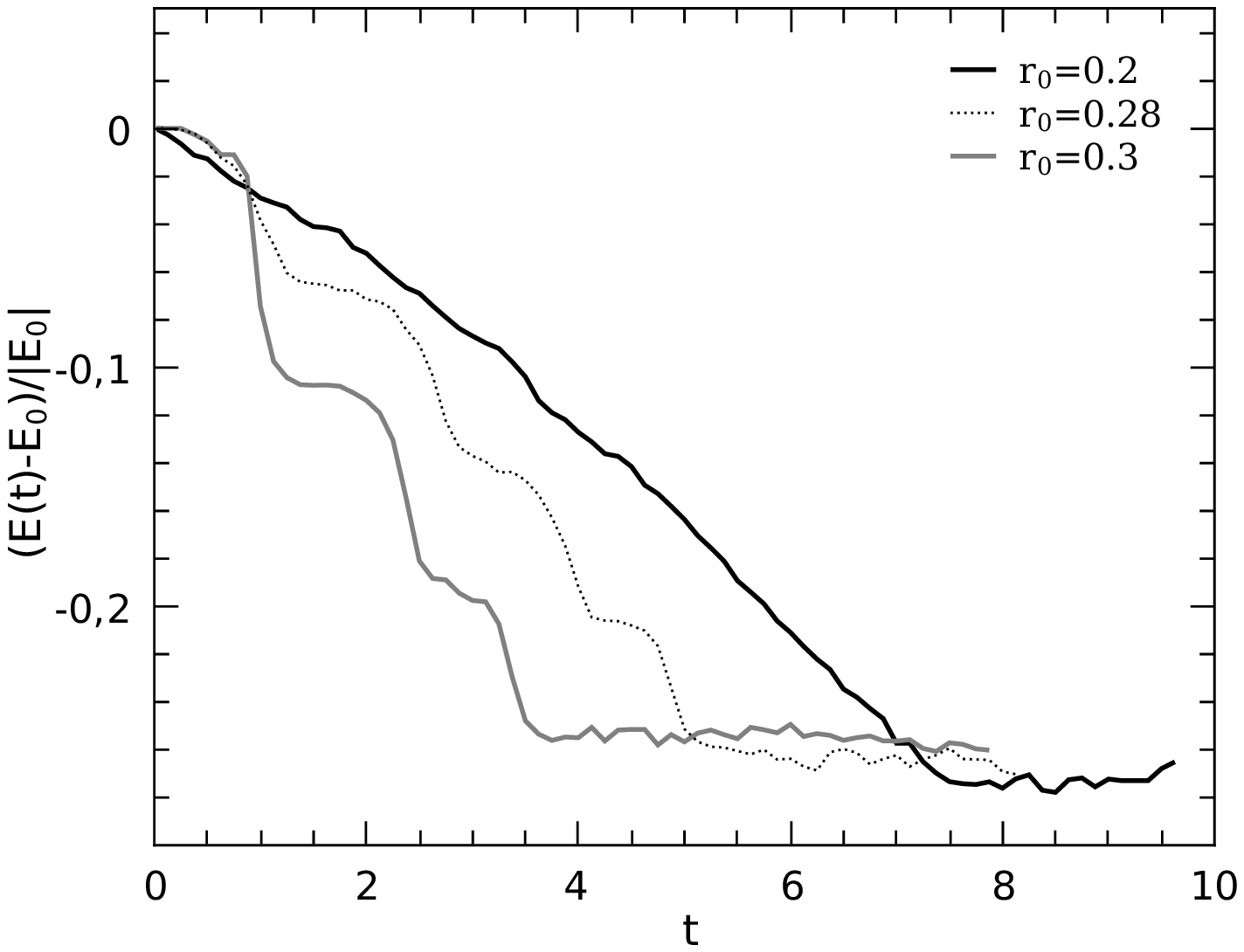}
\includegraphics[width=8cm]{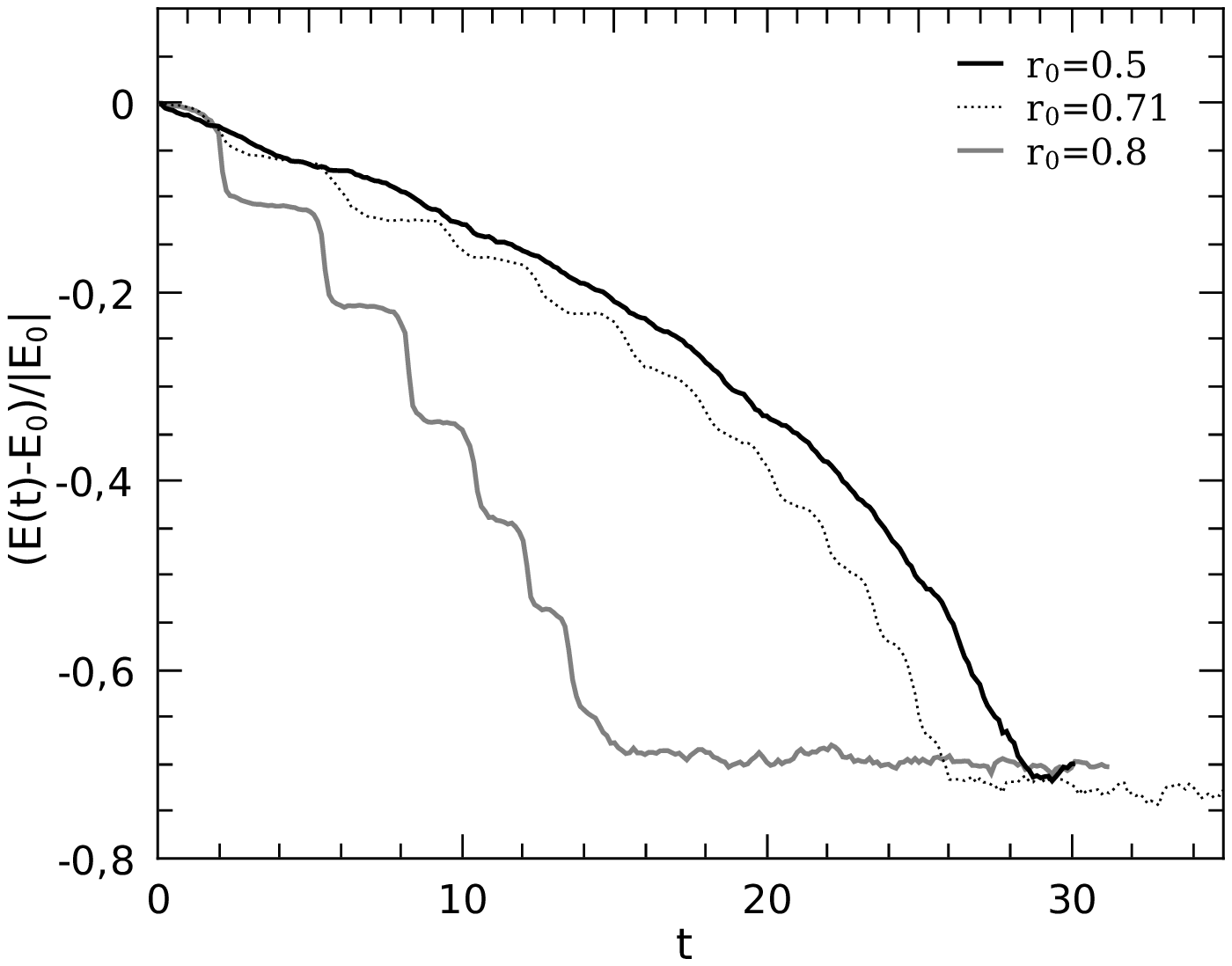}
\includegraphics[width=8cm]{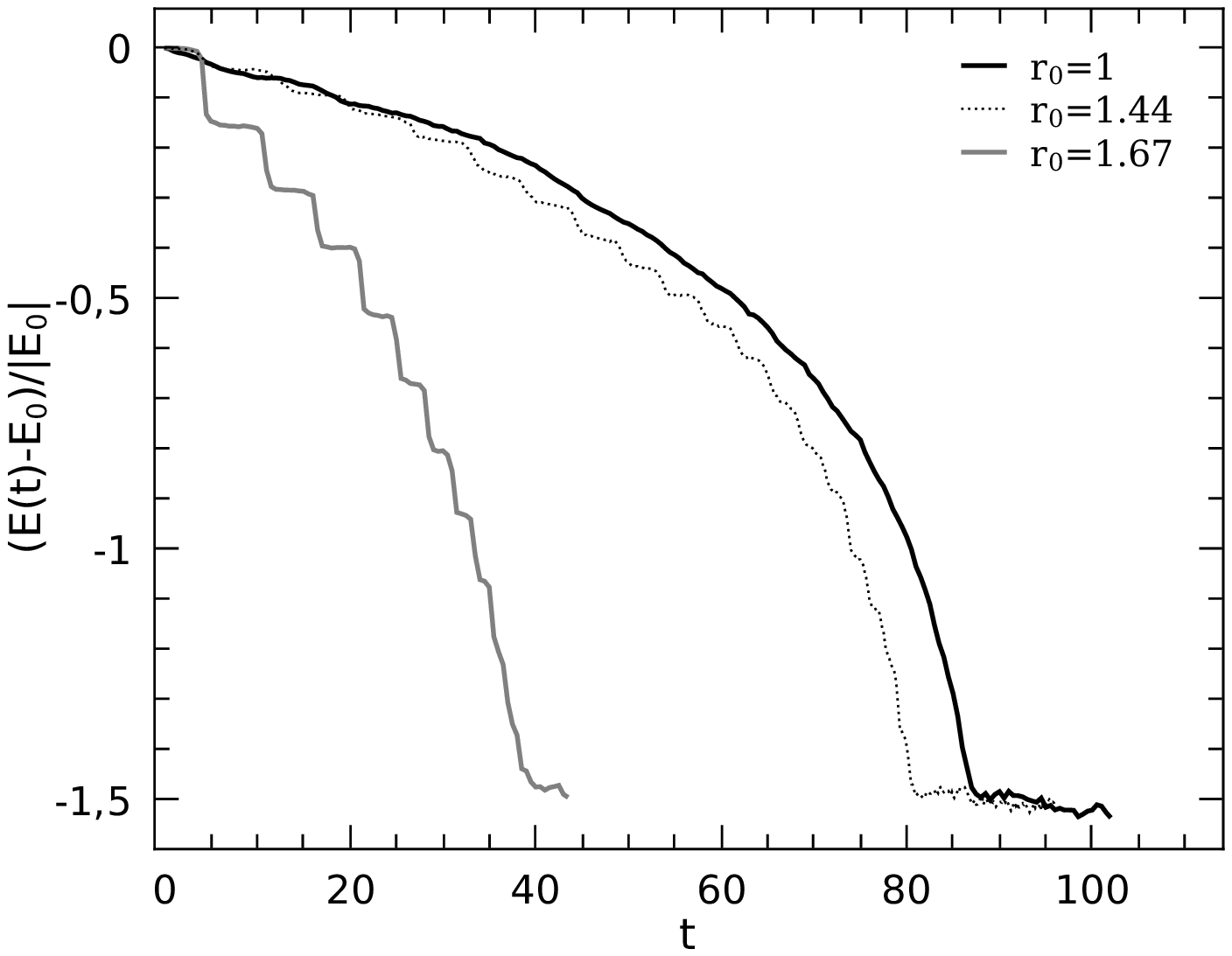}
\caption{Time evolution of the fractional variation of the test particle energy in the circular ($e=0$, solid black line), radial ($e=1$, dashed line) and an eccentric ($e=0.5$, grey line) cases of same initial energy, $E(0)$, in the $N$-body sampled, $\gamma=1$, galaxy.
The $r_0$ values refer to the initial distances of the test particle from the galactic center.}
\label{evstime}
\end{figure}

The role of the geometrical shape of the orbit is evident in Fig. \ref{evstime}, which shows the energy decay of the test mass for different initial eccentricities at fixed initial orbital energy.
As expected, circular ($e=0$) orbits decay slower than radial ($e=1$), while orbits with $e=0.5$ have decay time between these two extreme cases. We also note that, for higher initial orbital energies, the decay time of the $e=0.5$ orbit approaches that of the circular orbits, indicating a clear non-linearity of the decay time with $e$ in this high energy regime. 

Actually, the most important astrophysical parameter that can be inferred in this framework is the dynamical friction decay time, $\tau_{\rm{df}}$, which we define as the time needed to reduce the test particle orbital energy to $E(\tau_{\rm{df}})=\Phi(5\times 10^{-3}a)$. A correct evaluation of this time, which depends on both small and large scale characteristics of the galaxy where the test mass moves, as well on the test particle mass, is crucial in determining the actual role of dynamical friction in carrying matter toward the center of galaxies with the consequent, relevant, astrophysical implications.

\begin{figure}
\centering
\includegraphics[width=8cm]{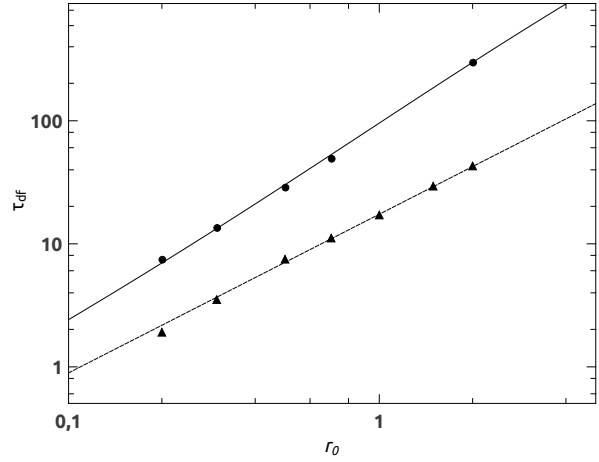}
\caption{Dynamical friction decay time vs. initial galactocentric distance for circular (filled circles) and radial (triangles) orbits in the $\gamma=1$ model.}
\label{tdfrc}
\end{figure}

\begin{figure}
\centering
\includegraphics[width=8cm]{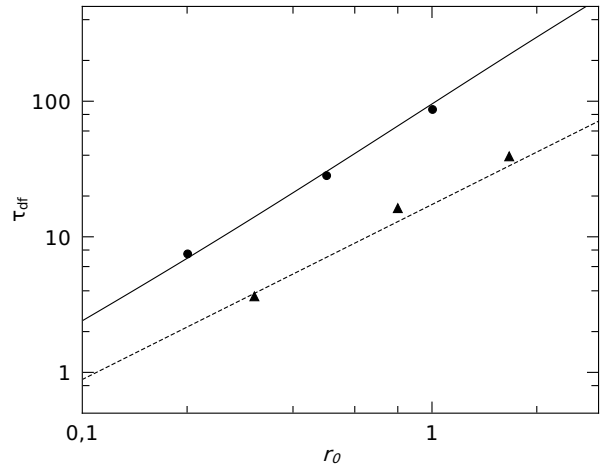}
\caption{As in Fig.\ref{tdfrc}, limiting the comparison to pair of orbits of same initial energy, $E(0)$.}
\label{sameen}
\end{figure}

Fig. \ref{tdfrc} shows the $\tau_{\rm{df}}$ dependence on the initial radial distance of circular and radial orbits  in the $\gamma=1$ model. The relations are two power laws 
with a slightly different slope. This is evident again in Fig. \ref{sameen}  where we compare $\tau_{\rm{df}}$ for circular and radial orbits with same initial energy.

Consider both circular and radial trajectories of same apocenter allow us to obtain an upper and lower limit, respectively, for decay time of any orbit at fixed position but different velocity. Simulating orbits with same initial energy, instead, we can study the efficiency of dynamical friction with respect to the shape of the orbit.

\begin{figure}
\centering
\includegraphics[width=8cm]{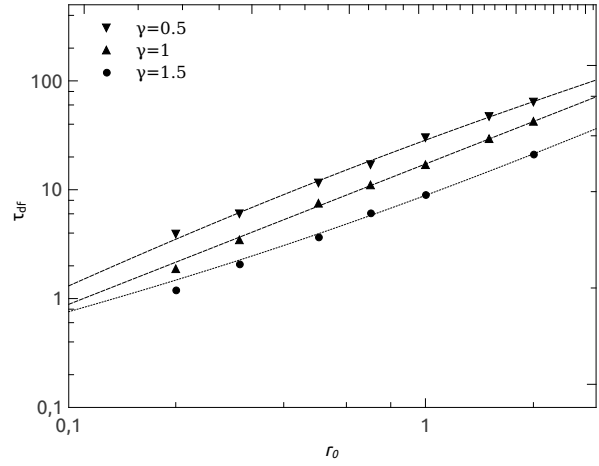}
\caption{Dynamical friction decay time vs initial galactocentric distance  for radial orbits in three different $\gamma$ models.}
\label{tdfgamma}
\end{figure}

The dynamical friction time depends, obviously, on the model considered: the steeper the density profile (large $\gamma$) the shorter the decay time.
This is clear in Fig. \ref{tdfgamma}, which shows how increasing $\gamma$ of a factor 3 (from $\gamma=1/2$ to $\gamma=3/2$), the decay time decreases by almost the same factor.

\subsection{Dynamical friction dependence on the test mass}

Beside the dependence from initial position, eccentricity and model, another important parameter that affect the dynamical friction effect is the mass
of the satellite. This dependence deserves some considerations. Actually, it is generally assumed a {\it direct}, linear, proportionality of the dynamical friction braking deceleration to the test mass, $M$.
This comes, in Eq. \ref{dfcorrect1}, by the contemporary assumption $m\ll M$ and $(b^2V^4)/(G^2(m+M)^2)\gg 1$. The opposite limit $(b^2V^4)/(G^2(m+M)^2)\ll 1$ would lead to an {\it inverse} linear proportionality. 
So it is logically inferred that performing the integrals in Eq. \ref{dfcorrect1} over the whole integration ranges lead to a dependence on $M^\alpha$ with $-1<\alpha<1$, even taking also into account a possible dependence of the integration limits on $m$ and $M$. 
\par\noindent We refer to Appendix B  for details. 

\begin{figure}
\centering
\includegraphics[width=8cm]{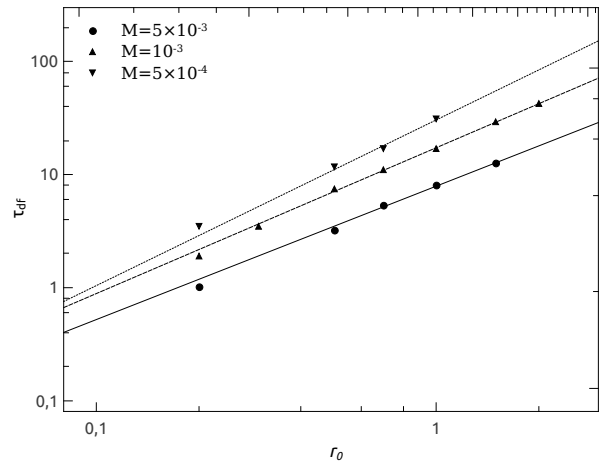}
\caption{Dynamical friction decay time vs initial galactocentric distance estimated from $N$-body simulations for radial orbits given three different values of the test particle mass, as labeled.}
\label{tdfmas}
\end{figure}

\begin{figure}
\centering
\includegraphics[width=8cm]{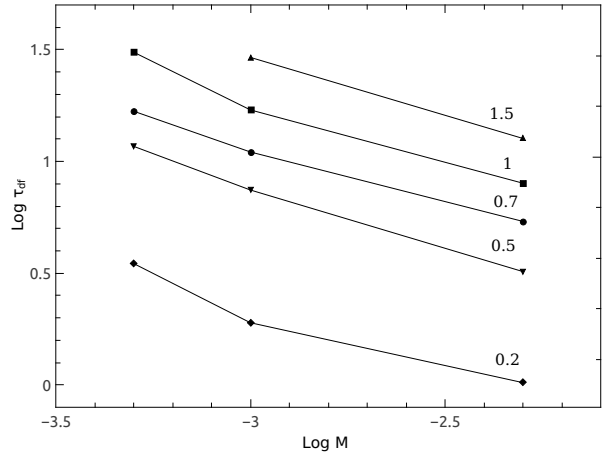}
\caption{Decay time as function of the test particle mass ($M=5\times 10^{-4},10^{-3},5\times 10^{-3}$) for initially radial orbits at different initial distances, as labeled.}
\label{mt}
\end{figure}

While it is confirmed that a higher mass of the test object leads to a shorter decay time (see Fig. \ref{tdfmas}) we see that, by varying the satellite mass in the range $[5\times 10^{-5},5\times 10^{-3}]$, the relation between $\tau_{\rm{df}}$ and $M$ is shallower:

\begin{equation}
\tau_{\rm{df}}\propto M^{-0.67\pm 0.1},
\end{equation}

as obtained by a least square fit to data of Fig. \ref{mt}, coming from direct $N$-body integrations and confirmed by the simplified one-body scheme.

\begin{figure}
\centering
\includegraphics[width=8cm]{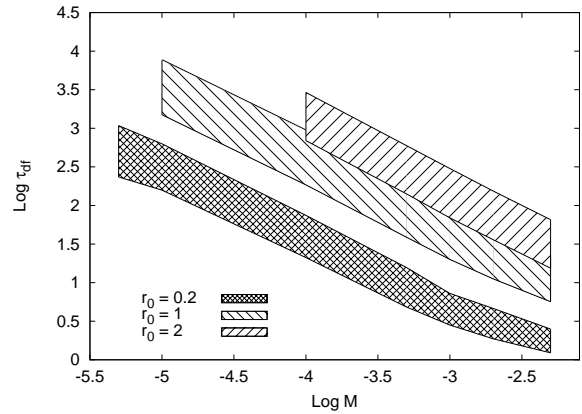}
\caption{Decay time in function of the test mass starting its motion from three different initial galactocentric distances in a $\gamma=1$ model. Each region is delimited by an upper line which refers to circular and a lower boundary defined by radial orbits. The decay times for all the other values of orbital eccentricity fall within these two boundaries.}
\label{distribution functiona}
\end{figure}

Assuming the minimum impact parameter independent of the test mass, we performed  semi-analytical simulations in a wide range of masses $[10^{-5}, 5\times 10^{-3}]$
setting the initial position $r_0$ to the values $0.2, 1$ and $2$ for initially circular and radial orbits, finding that the decay time-mass relation depends strongly on the starting position of the satellite (see Fig.\ref{distribution functiona}), as expected.

\subsection{A fitting formula for dynamical friction decay time}
\label{sec3}
A deep analysis of all the simulations done allowed us to obtain a useful analytical approximation to $\tau_{\rm{df}}$ in dependence on the relevant parameters, as

\begin{equation}
\tau_{\rm{df}}=\tau_0(1+g(e))(2-\gamma)M^{-0.67}r_0^{1.76},
\label{tdfit}
\end{equation}

where $\tau_0=0.2$ is an adimensional time constant and $g(e)$ is an adimensional function of the eccentricity:
\begin{equation}
g(e)=3.93(1-e).
\end{equation}

Eq. \ref{tdfit} is suited to give some useful astrophysical constraints. 
For example, for any given set of $\bar e$, $\bar{\gamma}$, $\bar{M}$ values, it gives 
the radius of the sphere containing all the test objects that, in a galaxy with a cusp $\gamma \geq \bar{\gamma}$, and having $e\leq \bar{e}$ and $M\geq \bar{M}$, have sunk to the galactic center within time $t$, as

\begin{equation}
r_{max}=2.5\left[\frac{t}{(1+g(\bar{e}))(2-\bar{\gamma})}\right]^{0.57}\bar{M}^{0.375}.
\end{equation}

\subsection{A straightforward application to a galactic satellite population sinking}

\begin{figure}
\centering
\includegraphics[width=8cm]{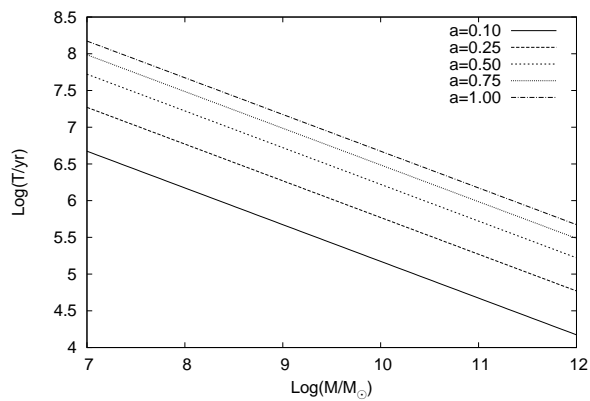}
\caption{The unit of time (Eq- \ref{PHt}) transformed into years, assuming galaxy masses ranging from $10^8M_\odot$ to $10^{12}M_\odot$ and for the length scale $a$ ranging from $0.1kpc$ to $1kpc$. This allows an easy rescaling of the decay time of Fig. \ref{distribution functiona} into a physical time for arbitrary choices of the pair $(M_G,a)$.}
\label{tunity}
\end{figure}

By mean of this formula and assuming a population of galaxy satellites (that may represent globular clusters in a galaxy) initially distributed following either the same $\gamma$ density law of the background stars or accordingly to a Plummer profile, we estimated the fraction to the total of satellites sunk to the center of the galaxy at different physical times (500 Myr, 1 Gyr and 13.7 Gyr), and synthesized some results in Tables \ref{frac0} and \ref{frac3/2}. 
It is clearly seen the fundamental role of the steepness of the galaxy density profile into the depletion of the satellite population, ass well as that the satellite mass.  
The cuspy, $\gamma=3/2$, galactic profile is able to erode around $40\%$ of the initial satellite population of masses larger than $M=10^5$ M$_\odot$ within 1 Gyr, assuming satellite moving on circular orbits, and up to $63\%$--$83\%$ of the initial population (the larger erosion for an initial satellite profile following the Plummer's law) in the case of radial ($e=1$) orbits. This erosion reduces to a $4\%$--$9\%$ of the initial satellite circular orbits and to $18\%$--$49\%$ of the initial satellite radial orbits, when the galaxy profile follows the, innermost flat, $\gamma=0$ profile (also here the percentages intervals refer to the satellites distributed as a $\gamma=0$ profile or as a Plummer's model). 
As a general conclusion, dynamical friction effect is maximized 
for massive satellites ($M/M_G\geq 10^{-6}$) of cuspy, massive and compact galaxies ($M_G\geq 10^{11}$ M$_\odot$, $a\leq 500$ pc) whose satellites systems evolve faster in a given physical time due to the $\propto a^{3/2} M_G^{-1/2}$ scaling of the time unit (see Fig. \ref{tunity}). 
In few Gyrs, such galaxies remain with a low abundant satellite population, having packed most of their mass (up to $90\%$, or more) into the galactic nuclear region.

\subsection{Massive object stalling in core galaxies}

\begin{figure}
\centering
\includegraphics[width=8cm]{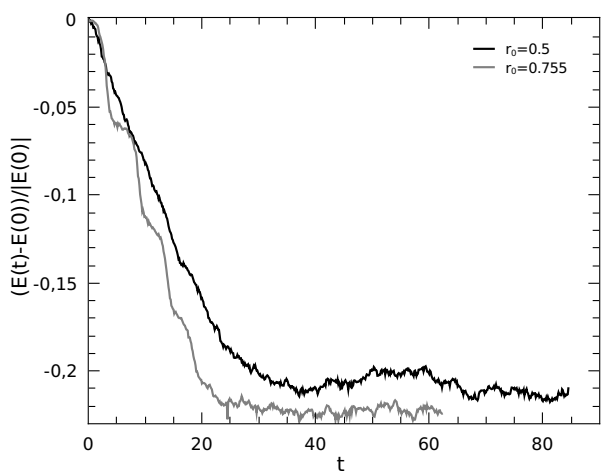}
\caption{Test mass energy loss for radial (dotted line) and circular (straight line) orbits in the case $\gamma=0$.}
\label{SEG0}
\end{figure}

The approximation formula given by Eq. \ref{tdfit} was obtained by fitting results of $N$-body integrations in cuspy density profiles. To check its application to cored models we performed two $N$-body simulations of the evolution of a radial and a circular orbit in a Dehnen model with $\gamma=0$. The orbits have same initial energy with the circular orbit starting at $r_0=0.5$. Fig. \ref{SEG0} reports the evolution of the test mass orbital energy in the two case studied.  

\begin{figure}
\centering
\includegraphics[width=8cm]{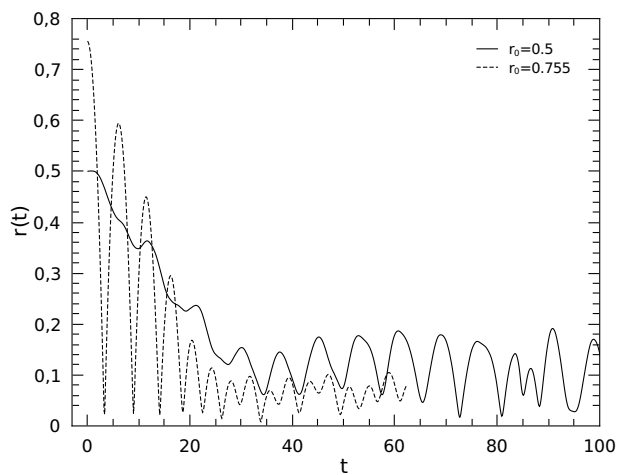}
\caption{Time evolution of the galactocentric distance of the radial (dotted line) and the circular (solid line) orbits (both obtained by N-body simulations) with the, labelled, initial distances from the center, in the $\gamma=0$ model. In the circular orbit, 
it is evident that at $t\sim 35$ the orbits become eccentric and the test particle almost stalls.}
\label{trG0}
\end{figure}

\begin{figure}
\centering
\includegraphics[width=8cm]{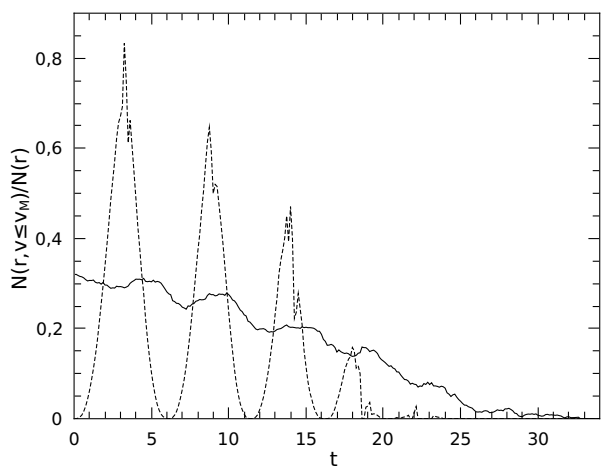}
\caption{The local fraction of field stars slower than the test mass as a function of time for a circular (straight line) and a radial (dotted line) orbit.}
\label{Nrat}
\end{figure}

We see that the extrapolation of Eq. \ref{tdfit} to the $\gamma=0$ case gives a decay time correct within $10\%$ for the radial orbit. On the other side, the $N$-body evolution of the circular orbit shows that the decay stops when the test particle galactocentric distance reduces to $r\lesssim 0.1$; then, the orbit \lq stalls\rq, in the sense that the test particle oscillates without appreciable further decay as indicated by Fig. \ref{trG0}. 
This orbit stalling in cored profiles was already found by previous authors \citep{kal72,readcole}; in particular, \cite{AntMer12} put on evidence 
that the stall is due to a lack of slow stars within the orbit size. 
Although it is not exactly true that dynamical friction is contributed by field stars slower than the decaying object, this interpretation is substantially correct as shown by Fig. \ref{Nrat}  
where the fraction (to the total) of stars slower than the decaying object and enclosed within its actual position is reported as function of time.

While in the radial case, the fraction of \lq slow\rq~increases when the 
test mass crosses the center of the system resulting into an enhancement of the dynamical friction effect which induces a progressive decay until the particle reaches the center of the system, in the circular case the fraction decrease continuously until $t\sim 30$, that is roughly the time at which the decay ends and the test mass reaches an almost steady eccentric orbit.

\begin{figure}
\centering
\includegraphics[width=8cm]{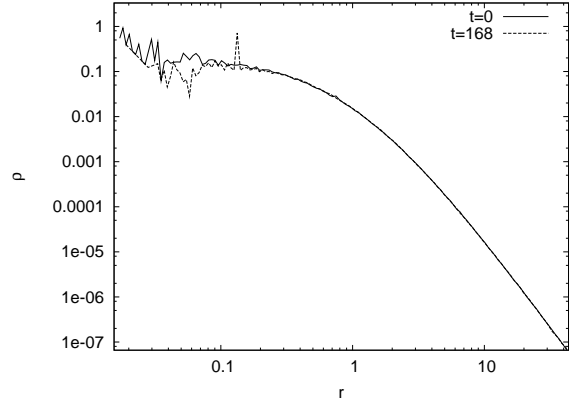}
\caption{Density profile of the backround distribution of particles at the beginning (solid line) and at the end of the simulation (dashed).}
\label{denscomp}
\end{figure}

Since the spatial distribution of background stars is not  significantly altered on all scales by the satellite motion, 
as it is shown in Fig.\ref{denscomp} where we compare the background density at the beginning and at the end of the simulation,
it is argued that the key parameter in the modes of braking is actually the variation in the number fraction of slow stars. 

Looking at the position at which stall begins, we found that the radius at which the dynamical friction action becomes negligible encloses a mass roughly equal to the test mass $M$, in agreement with the conclusion in \cite{Gual08}. Obviously, in flattened density core, this ``critical mass'' is reached at a greater radius with respect to cuspy profiles, enlarging the region of motion stalling. Of course, the stalling radius is smaller for centrally peaked profiles; for example, if $\gamma = 1$  it shrinks to $r\simeq 0.035$, as seen in Fig. \ref{tr3}.

\section{Indirect effects of a central black hole on the satellite decay}
\label{sec5}

It is well known that galaxies in a wide range of luminosities and Hubble types host at their center massive or even super massive black holes (SMBHs), whose masses range in the  $10^6-10^{10}M_\odot$ interval \citep{Antncc,UrPa}, influencing strongly the environment.

As an example of such an influence, \cite{AntMer12} noted that a hypothetical stellar-mass BH population would see enlarged significantly the time to reach the center of the Milky Way by the presence of the central SMBH.

Actually, the presence of an SMBH affects also larger space and time scale through, for instance, its indirect role on the dynamical friction efficiency.

To check this role, we performed some specific $N$-body simulations of the motion of a point-like object which starts on an initially radial orbit in a $\gamma=1$ sphere  sampled with $N=524,288$ particles and in presence of a central SMBH with mass $M_{BH}$.

In this framework, each background star has a mass $m_*\simeq 2\times 10^{-6}$.

We chose three different values for $M_{BH}$, namely $M_{BH}=M, 4M, 10M$, where the mass of the test object is set to $M=10^{-3}\gg m_*$.

Initial conditions for the test object are those of null initial velocity and of an initial position $\mathbf{r}_0=(x_0>0,0,0)$ such that the initial orbital energy of the test object is the same in the three cases, $E_0=-5\times 10^{-4}$. This choice leads to about the same speed at the closest approach of the test particle to the center, condition needed to appreciate differences in the decay as mainly due to the presence of the black hole.

\begin{figure}
\centering
\includegraphics[width=8cm]{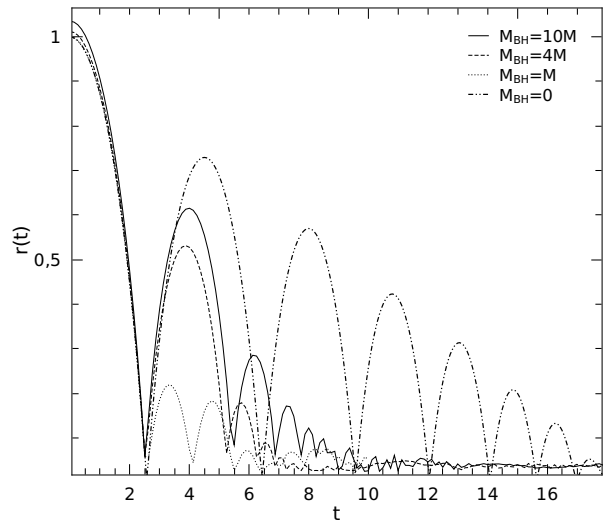}
\caption{Test mass, $M$, orbital decay in presence of a central black hole, whose mass, $M_{BH}$, is labelled. The galaxy is modelled as a Hernquist sphere and the test mass motion computed in the complete N-body framework.}
\label{BHdec}
\end{figure}

\begin{figure}
\centering
\includegraphics[width=8cm]{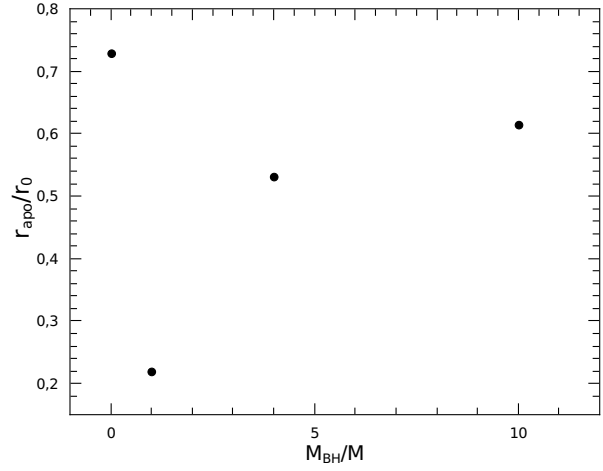}
\caption{The apocenter distance after the first oscillation through the galactic center of the test mass in presence of a central massive black hole of mass $M_{BH}$.}
\label{apocenters}
\end{figure}

\begin{figure}
\centering
\includegraphics[width=8cm]{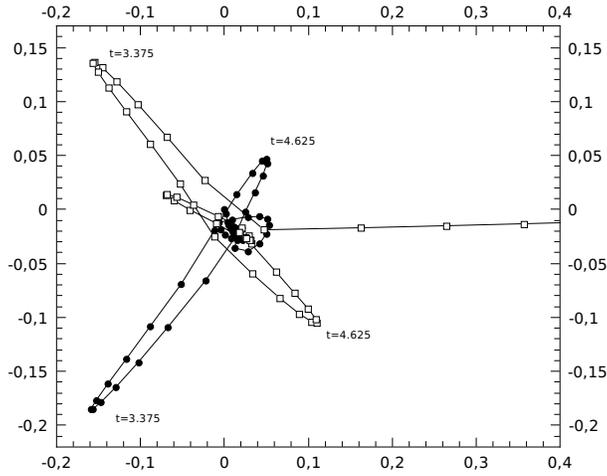}
\caption{The trajectories of the radially falling test mass (empty squares) and of the perturbed central BH (filled squares), in the case of equal mass. Some of the apocenter positions are labeled with their times.}
\label{BHtrajectory}
\end{figure}

\begin{figure}
\centering
\includegraphics[width=8cm]{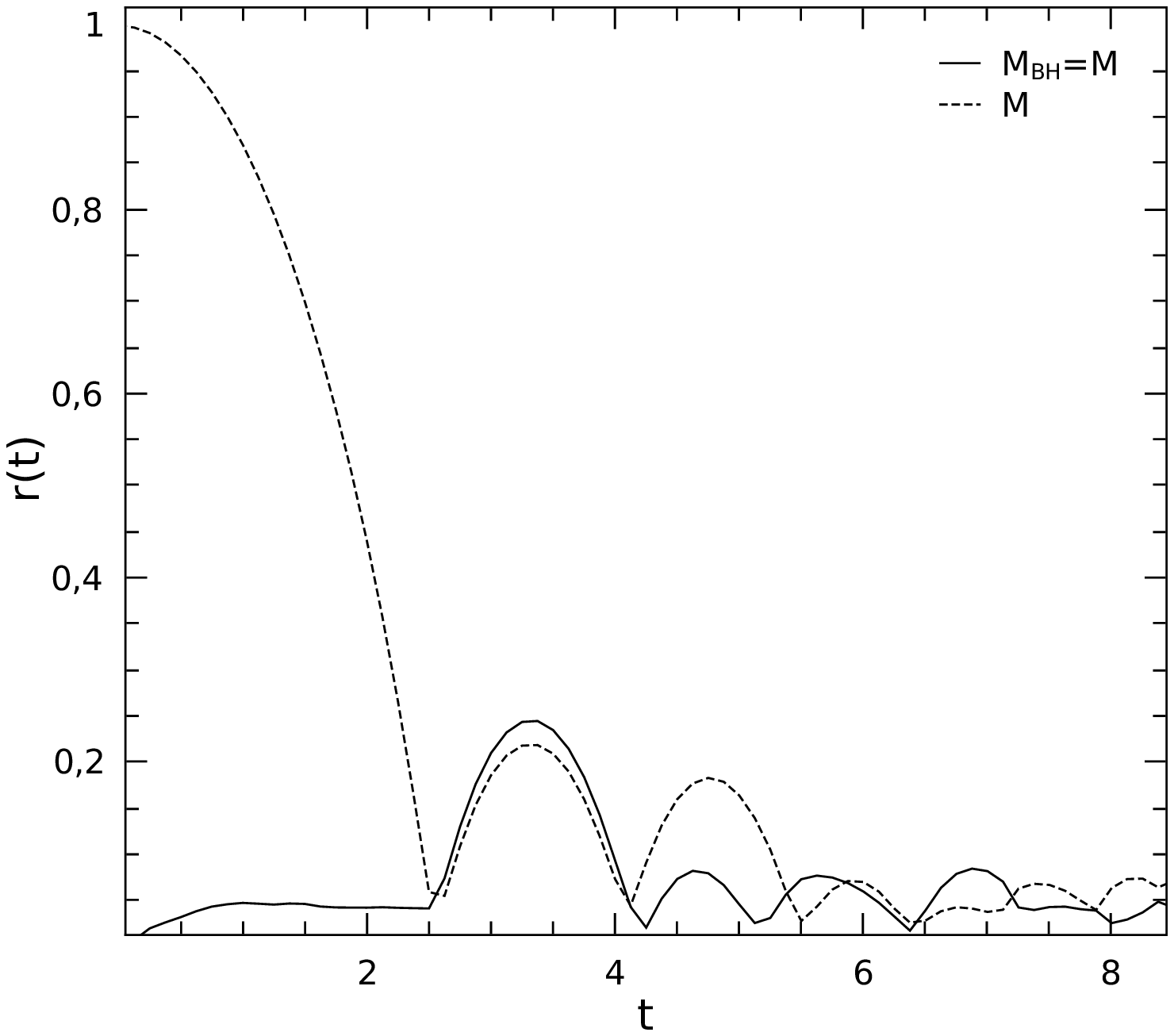}
\includegraphics[width=8cm]{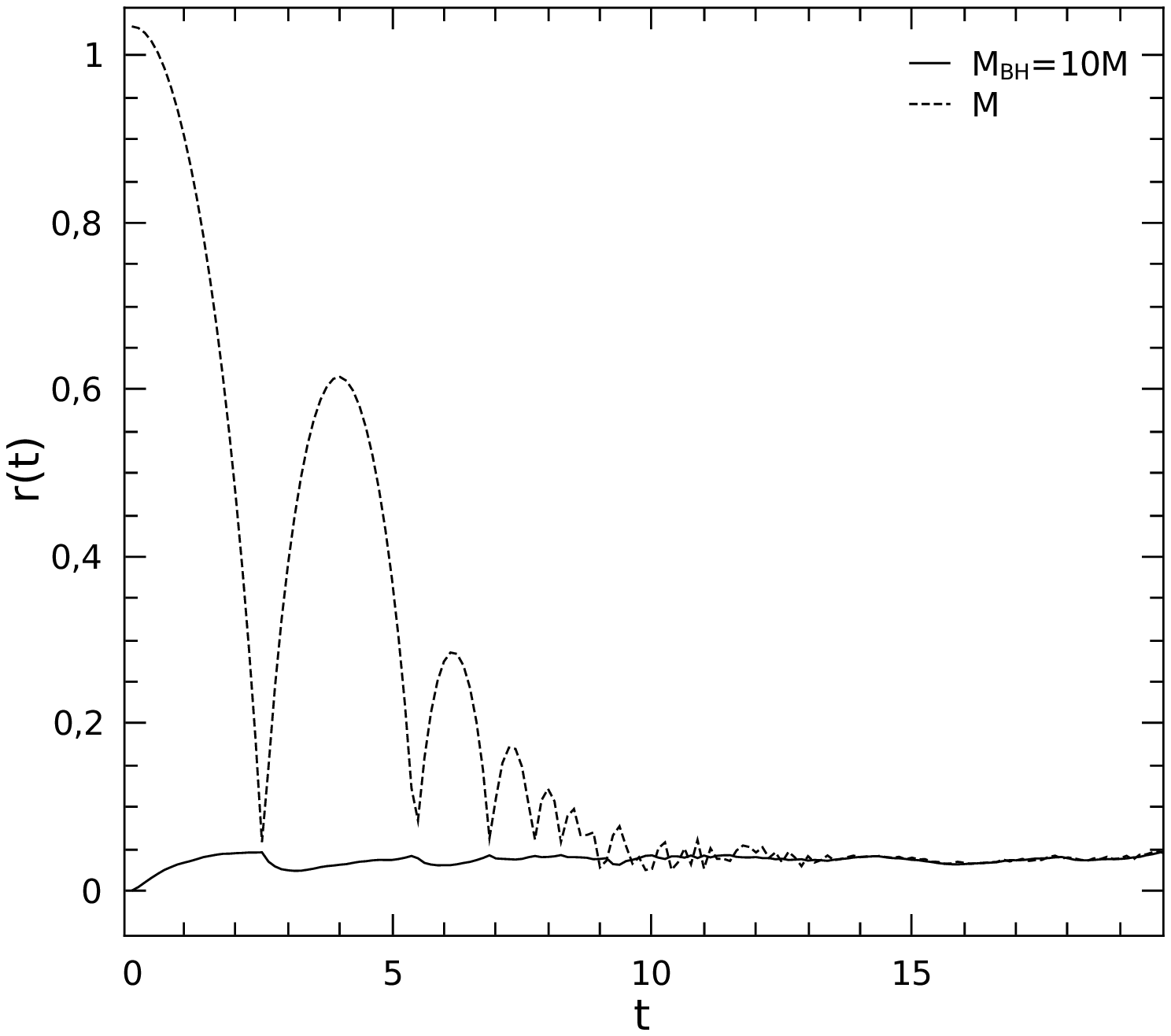}
\caption{Test mass (\textit{dotted line}) and BH (\textit{straight line}) galactocentric distances vs. time in the case $M_{BH}=M$ (\textit{upper panel}) and $M_{BH}=10M$ (\textit{lower panel}).}
\label{orbcmp}
\end{figure}

\begin{figure}
\centering
\includegraphics[width=8cm]{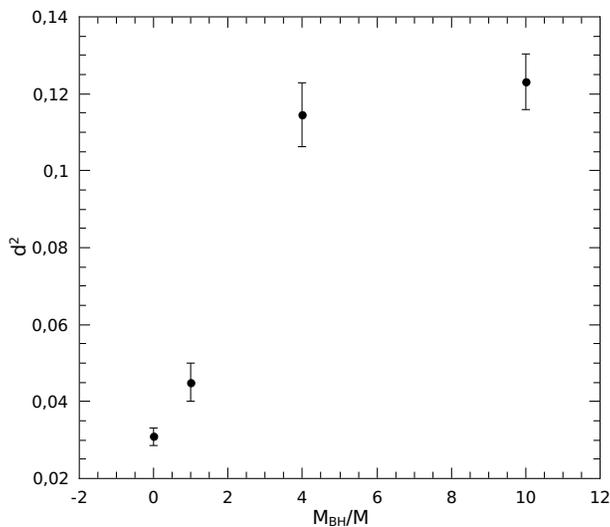}
\caption{Cumulative standard deviation from the unperturbed radial motion of the test mass $M$ as a function of the BH mass;  which shows the standard deviation of the distance of the test mass from the direction of unperturbed radial motion evaluated over the whole orbital evolution of the test mass until its total decay, as function of $M_{BH}$.}
\label{departure}
\end{figure}

The time evolution of the test object distance to the galactic center, shown in Fig. \ref{BHdec}, indicates that the presence of a SMBH does affect the dynamical friction decay time.
More massive BH determines a longer decay time of the infalling object. It should not surprise that the behaviour of $r(t)$ in the case of abscence of SMBH is more similar to the behaviour in the case of the most massive SMBH considered. This is due to that the apocentric distance reached after the first crossing through the center is much more similar  in these two extreme cases than in the others because the very massive BH after the close encounter with the test particle gains just a small velocity. Less massive SMBHs, on the other hand, move more and the test mass apocenter reduces consequently, making it moving in an innermost region where the galactic dynamical friction effect is larger. This is made clear by Figs. \ref{apocenters}, \ref{BHtrajectory} and 
\ref{orbcmp}. This effect dominates on the other, opposite, effect of deviation from the unperturbed radial trajectory as quantified in Fig. \ref{departure}.  
This figure shows a very similar time for the closest approach to the galactic center (and so to the SMBH therein) in all the cases studied ($t\simeq 2.5$), consequence of the same value of initial orbital energy. After this closest approach, the time evolution of the distance to the center is quite different and differences cumulate over the following closest approaches. 
 
The effect of the interaction BH-test mass is clearly shown in Fig. \ref{BHtrajectory},  which draws the trajectories (labelled with times) of the test mass and of the SMBHs in the case $M_{BH}/M=1$ with the clear departure of the central BH from its initial central position.

The effects induced by the presence of a central black hole on the motion of the satellite can change significantly the time needed to carry the satellite  toward the center of the system. Here, we have shown that the decay time is minimum when the sinking satellite and the central black hole have about same mass, while when the black hole mass exceeds several times the mass of the infalling object the decay time tends to the same value estimated in asbcence of a central black hole. This implies that results presented in Sect.\ref{sec3} are still valid whenever the central body is significantly more massive than the incoming satellite. This is often the case of real galaxies, at least for galaxies more massive than $\sim 10^{10}$M$_\odot$ \citep{scot}. In these massive hosts, the fitting formula given in Eq.\ref{tdfit} represents a valid way to measure the amount of mass deposited in time within the central region of a galaxy also if it hosts a central, massive, black hole.

\section{Conclusions}
In this paper we studied dynamical friction in cuspy density profiles of spherical (E0) galaxies, both on a theoretical and a numerical point of view.

The main results are here summarized:
\begin{enumerate}
\item the classic \cite{Cha43I} formula in its local approximation does not work in the central region of a cuspy distribution, because it diverges at the center and overestimates the actual dynamical friction in the vicinity of the density singularity;
\item an alternative, semi-analytic expression for the dynamical friction formula (Eq. \ref{interp}) which is finite at center of density diverging galaxies (as mathematically shown in this paper Appendix A in the case of the family of \citet{Deh93} $\gamma$ models) and smoothly connected to the usual local approximation is given and discussed;
\item the free parameters in the semi-analytic formula are tuned via comparison with high precision $N$-body simulations of massive object decay in a self consistent particle representation of the cuspy host galaxy (Sect. 3); the best values of the minimum impact parameter is systematically larger for circular ($e=0$) orbits than for radial ($e=1$);
\item an extensive set of orbits of different initial eccentricities for a massive test object in the $N$-body representation of the parent galaxy has been computed,
showing both a good agreement with the semi-analityc formula as shown by Figs. \ref{enratio}--\ref{tr5};
\item for any given initial orbital energy, the decay times of orbits of different eccentricities range within the interval defined by radial (shortest) and circular (longest) case;
\item the ratio of the radial to circular decay times in the case of the $\gamma=1$ density slope is about $1/2$;
\item global approximation formulas for the dynamical friction decay time in function of the relevant structural parameters are obtained, which show clearly how dynamical friction is maximized in massive host galaxies with a steeper central density profile, for higher eccentricity orbits of massive satellites;
\item as an example, our Milky Way, if represented in its central region as a moderate cuspy density ($\gamma = 1/2$) should have lost, in a Hubble time,  about $75 \%$ of the initial population of massive ($\geq 10^5$ M$_\odot$) globular clusters, decayed into the innermost region;
\item the dynamical friction decay of test objects is altered significantly by the presence of a central massive black hole \textit{if it has a mass comparable to the satellite mass}; the decay time of initially radial orbits is an increasing function of $M_{BH}$;
\item on the other hand, when the central BH has a mass significantly greater than the satellite mass, the decay time is well estimated by our general formulas;
\item the dynamical friction time, $\tau_{df}$, depends on the test object mass in a non-trivial manner, which is different from the usually adopted inverse linearity,  $\tau_{df}\propto M^{-1}$, resulting  $\tau_{df}\propto M^{-0.67}$, instead.
\end{enumerate} 

\section{Acknowledgements}
We thank the Aspen Center for Physics for the hospitality during the development of part of this work. Thanks are also due to M. Spera for his collaborative help in the use of HiGPUs code and to F. Antonini for discussions on the role of massive black holes for the main topic of this paper. Acknowledgements are due also to an anonymous referee whose suggestions allowwed us to improve the quality of the paper.

\section*{Appendix A}

In this paper we use as self consistent models of spherical, cuspy galaxies the distribution functions that represent the 
so called \textit{Dehnen's} (or \textit{gamma}) models \citep{Deh93} that are the 3-parameters density distributions following the laws 

\begin{equation}
\rho(r)=\frac{(3-\gamma)M}{4\pi a^3}\frac{1}{(r/a)^\gamma(1+r/a)^{4-\gamma}},
\label{rhoD}
\end{equation}

where $0\leq \gamma \leq 3$ gives the slope of the centrally diverging (if $\gamma >0$) profile, $a$ is the length scale,and $M$ is the total mass of the model.
The case $\gamma=0$ corresponds to a central core, where density flattens. The cases $\gamma=1$, and $\gamma=2$ correspond to the classic \citet{Hern90} and \citet{Jaffe83} models, respectively.

The density profile $\rho(r)$ of Eq.\ref{rhoD} can be expressed as a function of the potential $\Psi(r)$
so that it is possible to apply the \citet{Edd16} inversion formula to obtain the unknown distribution function $f(\mathcal{E})$ as:

\begin{equation}
f(\mathcal{E})=\frac{(3-\gamma)}{2(2\pi^2 GMa)^{3/2}}\int_0^\mathcal{E} \! \! \frac{(1-x)^2[\gamma +2x +(4-\gamma)x^2]}{x^{4-\gamma}\sqrt{\mathcal{E}-\Psi}}d\Psi,
\label{disfunc}
\end{equation}

where
\begin{equation}
x\equiv x(\Psi)= \left\{
\begin{array}{lcl}
e^{-\Psi} & \qquad & \gamma=2 \\
 & & \\
\left[1-(2-\gamma)\Psi\right]^{1/(2-\gamma)} & \qquad & \gamma\neq2 \, .
\end{array}
\right.
\end{equation}

If $(2-\gamma)^{-1}$ is integer or half-integer, the integral in Eq.\ref{disfunc} can be calculated in terms of linear combination of hypergeometric series, easily reduced to elementary functions \citep{GradR}. 

Due to that cusps steeper than $\gamma = 2$ are not observed in real galaxies, we limited to consider the cases $\gamma=0,\ 1/2,\ 1,\ 4/3,\ 3/2,\ 7/4,\ 2$, which are all values leading to $(2-\gamma)^{-1}$ integer or half-integer leading to analytic expressions for $f(\mathcal{E})$, but $\gamma=1/2$ which deserves a numerical integration to get $f(\mathcal{E})$ which was later fitted in a way to have this general expression for the isotropic distribution function:

\begin{equation}
f_\gamma(\mathcal{E})=  
\frac{M}{(GMa)^{3/2}} \frac{A_\gamma}{(\Psi(0)-\mathcal{E})^{(6-\gamma)/(2(2-\gamma))}}
					    \left[g_\gamma(\mathcal{E})+B_\gamma\sqrt{\mathcal{E}}\sqrt{\psi(0)-\mathcal{E
}}\left(\sum\limits_{i=0}^{(2+\gamma)/(2-\gamma)}					   
b_i\mathcal{E}^i\right)\right],
\label{distribution function}
\end{equation}

where $g_\gamma(\mathcal{E})$ is

\begin{equation}
g_\gamma(\mathcal{E})=
\begin{cases}
(3-4\mathcal{E})\sqrt{2\mathcal{E}}\sqrt{\Psi(0)-\mathcal{E}}-3\sqrt{(\Psi(0)-\mathcal{E
})^3} \log\left(\frac{1+\sqrt{2\mathcal{E}}}{\sqrt{1-2\mathcal{E}}}\right), & \gamma=0 \\
3\arcsin{\sqrt{\mathcal{E}}}, & \gamma=1 \\
54675\sqrt{2}\arcsin{\sqrt{\frac{2\mathcal{E}}{3}}}-450\sqrt 6 (3-2\mathcal E)^{9/2}\log\left({\frac{3+2\sqrt{6\mathcal E} +2\mathcal E}{3-2\mathcal E}}\right), & \gamma=4/3 \\
3(3+32\mathcal E-8\mathcal E^2)\arcsin{\sqrt{\frac{\mathcal E}{2}}}, & \gamma=3/2 \\
-33633600(83-512\mathcal E+192\mathcal E^2-32\mathcal E^3+2\mathcal E^4)\arcsin{\frac{\sqrt\mathcal E}{2}}. & \gamma=7/4\\
\end{cases}
\label{eq:qw}
\end{equation}

and the values of $A_\gamma$ and of $b_i$ are reported in Table A1.

Finally, the $\gamma=2$ case (Jaffe's model) has a formal expression that is not easily reduced into the form of Eq. \ref{distribution function}; as known \citep{Jaffe83} it is given by 

\begin{equation}
f(\mathcal{E})=\frac{M}{2\pi^3(GMa)^{3/2}}\left[ F_-\left(\sqrt{2\mathcal{E}}\right)-\sqrt{2}F_+\left(\sqrt{2\mathcal{E}}\right)-\sqrt{2} F_-\left(\sqrt{2\mathcal{E}}\right)+F_+\left(\sqrt{2\mathcal{E}}\right)\right]
\end{equation}

where

\begin{equation}
F_\pm (\eta) = e^{\mp x^2}\int_0^x e^{\pm \eta^{2}}d\eta.
\end{equation}

\subsection*{The convergence of the dynamical friction integral}

We study here the convergence of the dynamical friction integral in Eq. \ref{dfcorrect1}, which is an improper integral in the case of the a cuspy matter density distributions such as the case of the family of the \lq gamma\rq~laws given by Eq. \ref{rhoD}. 
Only when $\gamma =0$ (which means a central core) the dynamical friction integral is not singular, while it is for any $\gamma >0$. In these cases, the adoption of the distribution functions in their limit for high binding energies as expressed by Eq. \ref{eq:qw} leads to a dynamical friction integral which in a neighbourhood of the origin of the phase-space (that is for a slow motion around the galactic center) assumes the form:



\begin{equation}
\frac{\rm{d}{\mathbf v}_{\it M}}{\rm{d}t} = -A\int_{b_{min}}^{b_{max}}\! \! \int 
\frac{1}{2}\left[\frac{v^2}{(GM)/a} + (r/(r+a))^2\right]
\frac{\mathrm{V}b}{1+b^2{\mathrm V}^4G^{-2}(m+M)^{-2}}
{\mathbf V}\rm{d}^3\mathbf{v}_{\it m}\rm{d}b,
\label{}
\end{equation}

for $\gamma = 0$, and

\begin{equation}
\frac{\rm{d}{\mathbf v}_{\it M}}{\rm{d}t} = -A\int_{b_{min}}^{b_{max}}\! \! \int 
\! \left[\frac{1}{2}\frac{v^2}{(GM)/a}
+\frac{1}{2-\gamma}\left(\frac{r}{r+a}\right)^{2-\gamma}
\right]^{-(6-\gamma)/(2(2-\gamma))}
\frac{\mathrm{V}b}{1+b^2{\mathrm V}^4G^{-2}(m+M)^{-2}}
{\mathbf V}\rm{d}^3\mathbf{v}_{\it m}\rm{d}b,
\label{e1}
\end{equation}

for $0<\gamma<2$, and

\begin{equation}
\frac{\rm{d}{\mathbf v}_{\it M}}{\rm{d}t} = -A\int_{b_{min}}^{b_{max}}\! \! \int 
e^{-v^2/((GM)/a}\left(\frac{r}{r+a}\right)^{-2}
\frac{\mathrm{V}b}{1+b^2{\mathrm V}^4G^{-2}(m+M)^{-2}}
{\mathbf V}\rm{d}^3\mathbf{v}_{\it m}\rm{d}b,
\label{e2}
\end{equation}

for $\gamma=2$, and


\begin{equation}
\frac{\rm{d}{\mathbf v}_M}{\rm{d}t}\! =\! -A\int_{b_{min}}^{b_{max}} \! \!
\! \! \int \! \! 
\left\{\frac{1}{2-\gamma}\left[1-\left(\frac{r}{r+a}\right)^{2-\gamma}\right]- \! \! \frac{v^2}{2b}\right\}^{[6-\gamma]/[2(2-\gamma)]}
\frac{\mathrm{V}b}{1+b^2{\mathrm V}^4G^{-2}(m+M)^{-2}}
{\mathbf V}\rm{d}^3\mathbf{v}_{\it m}\rm{d}b,
\label{e3}
\end{equation}

for $2<\gamma<3$.

The convergence of the above improper integrals can be studied by analysing the properties of the integrands (which we call $I_1, I_2$ and $I_3$, respectively) for $r/a$ and $(1/2)v^2/(GM)/a$ going contemporarily to zero (i.e. with the same order), introducing the auxiliary infinitesimal variable $x\equiv r/a=v^2/(2b)$. This way, it is easily seen that the four integrands behave, for $x\ll 1$, as:

\begin{align}
I_1 &\approx x\left(1+\frac{1}{2}x\right)^{-1}, & \gamma=0,\\ 
I_2 &\approx  x^2\left[x(1+x^{1-\gamma})\right]^{-(6-\gamma)/(2(2-\gamma))}, &0<\gamma<2,\\
I_3 &\approx  e^{-x}, &\gamma=2,\\
I_4 &\approx  x^{(10-\gamma)/2}, &2<\gamma<3
\end{align}

In the case of Eq. \ref{e1}: if $0<\gamma\leq 1$ the behaviour is $x^{(2-3\gamma)/(2(2-\gamma))}$ whose exponent is $\geq -1/2$, implying the integral convergence; if $1<\gamma<2$ the behaviour is $x^{-(2-\gamma)/2}$ whose exponent is $\geq -1$, which again guarantees convergence. In the cases of Eq. \ref{e2} and Eq. \ref{e3} 
the limits are again finite, different from zero when $\gamma=2$ and equal to zero for $2<\gamma<3$. Note that this latter case has not always an acceptable physical meaning because it may give negative values for the distribution function around the origin of the phase space.

\section*{Appendix B}
While the hypothesis of dynamical friction as cumulative effect of multiple hyperbolic encounters implies a growth of its effect at increasing values of $M$, the integral in Eq. \ref{dfcorrect1} is such that the final dependence on $M$ may be different than a simple proportionality to $M$, although in the limit $m\ll M$.

Actually, the expression for dynamical friction given by Eq. \ref{interp} contains two additive terms. The local term (Eq. \ref{local}) has an explicit, dominant linear dependence on $m+M$ and another, weaker, dependence through the Coulomb's logarithm (essentially a $\ln (b_{max}/b_{min})$ dependence). The central term (Eq. \ref{dfcen1}) has an inversely linear dependence on $m+M$ in the multiplicative factor of the integral and depends on $M$ also in the integrand and in the integration limits. 
Applying the Leibnitz's formula for the integral differentiation we have three terms in the derivative respect to $M$:

\begin{equation} 
\label{Mderivative}
\begin{split}
\frac{1}{4\pi m} & \frac{d}{dM} \left(\frac{\mathrm{d} \bf{v}_{\it M}}{\mathrm{d}t}\right)_{\mathrm{cen}} = 
-\int_{b_{min}}^{b_{max}} \! \! \! \int 
\! f(b,{\bf{v}_{\it m}})\frac{d}{dM} 
\frac{V}{(m+M)\left[1+b^2 {\mathrm V}^4G^{-2}(m+M)^{-2}\right]}  \mathbf{V} \mathrm{d}^3 \mathbf{v_{\it m}} b \mathrm{d}b+\\
& -\frac{db_{max}}{dM}\int \! f(b_{max},{\bf{v}_{\it m}}) 
\frac{V}{(m+M)\left[1+b^2 {\mathrm V}^4G^{-2}(m+M)^{-2}\right]}  \mathbf{V} b_{max}\mathrm{d}^3 \mathbf{v_{\it m}}+\\
& +\frac{db_{min}}{dM}\int \! f(b_{min},{\bf{v}_{\it m}}) 
\frac{V}{(m+M)\left[1+b^2 {\mathrm V}^4G^{-2}(m+M)^{-2}\right]}  \mathbf{V} b_{min}\mathrm{d}^3 \mathbf{v_{\it m}}.
\end{split}
\end{equation}

The first term results to be

\begin{equation} \label{Mder1}
\frac{1}{(m+M)^2} \int_{b_{min}}^{b_{max}} \! \! \! \int 
\! f(b,\mathbf{v}_m)
\frac{1-b^2{\mathrm V}^4G^{-2}(m+M)^{-2}}{\left[1+b^2{\mathrm V}^4G^{-2}(m+M)^{-2}\right]^2}
V\mathbf{V} \mathrm{d}^3 \mathbf{v}_m b \mathrm{d}b,
\end{equation}

which tends to a mass independent value (i.e. linearity of dynamical friction deceleration in $m+M$) only in the weak encounter regime 

\begin{equation}
\frac{b^2 {V}^4}{G^2(m+M)^2} \gg 1,
\end{equation}

while in the opposite (strong encounter) regime

\begin{equation}
\frac{b^2 {V}^4}{G^2(m+M)^2}  \ll 1, \\
\end{equation}

it shows an inverse quadratic dependence on $m+M$ (lighter test masses would be more strongly decelerated).

Regarding the other two terms in Eq. \ref{Mderivative} the first is usually set to 0 by the assumption of $b_{max}$ as the, fixed, characteristic length size of the system, while the second depends on the choice for $b_{min}$. 
For a generic dependence of $b_{min}$ on $M$, the dependence on $M$ through the explicit derivative of $b_{min}$ respect to $M$ is modulated by the dependence on $b_{min}$ in the integrand. If we impose to $b_{min}$ the logical constraint to be large enough to allow 2-body hyperbolic encounters, only, something like $b_{min}=G(m+M)/v_\infty^2$ (where $v_\infty$ is the speed of the free test mass) is obtained, whose derivative respect to $M$ is $G/v_\infty^2$. Hence, the second term has the likely dominant dependence on $M$ in its explicit part in the integrand and, thus neglecting the dependence on $M$ through $f(b_{min},\mathbf{v}_m)$, we have that the regime:

\begin{equation}
\frac{b^2 {V}^4}{G^2(m+M)^2} \gg 1, \\
\end{equation}

gives a direct linear dependence on $m+M$ (i.e. quadratic in dynamical friction), 
while in the opposite regime 

\begin{equation}
\frac{b^2 {V}^4}{G^2(m+M)^2} \ll 1, \\
\end{equation}

the dependence is inversely linear in $m+M$ (i.e. a logarithmic dependence of dynamical friction on $m+M$).

From what we said above, also assuming that dynamical friction is mainly contributed by the cumulation of many weak encounters, its dependence on mass is not simply linear in $m+M$ but it is altered by an additive $\ln (m+M)$ dependence whose amplitude is modulated by the degree of spatial divergence of the distribution function.
In any case, the expected dynamical friction dependence on $m+M$ is something like $\propto (m+M)^\alpha$, with $0<\alpha<1$.

\bibliography{ASCDbib}

\clearpage

\clearpage

\clearpage 

\begin{table}
\caption{}
\centering{The critical radius and the radius that encloses $10\%$ of the total mass. The last column reports the relative variation between the two.}
\begin{center}
\begin{tabular}{c c c c}
\hline
\hline
   $\gamma$ & $r_{cr}$  & $r(0.1)$ & $\Delta r/r ~\%$ \\ 
\hline
     $0.5$  &   $0.7$   &  $0.661$  &    $0.059$ \\
     $1.0$  &   $0.5$   &  $0.463$  &    $0.080$ \\
     $1.5$  &   $0.3$   &  $0.275$  &    $0.091$ \\
\hline
\end{tabular}
\end{center}
\label{rcr}
\end{table}


\begin{table}
\caption{}
\centering{Parameters defining the $N$-body $\gamma=1/2$ simulations.}
\begin{center}
\begin{tabular}{c c c c c c}
\hline
\hline
$r_0$ & $e=0$ & $e=0.50$ & \multicolumn{3}{c}{$e=1$} \\
\hline
$ $  &  $M=10^{-3}$ & $M=10^{-3}$ & $M=5\times 10^{-4}$ & $M=10^{-3}$ & $M=5\times 10^{-3}$ \\
\hline
$0.2$  & $\checkmark$ & $-$ &$-$ & $\checkmark$ & $-$ \\
$0.3$  & $-$ & $-$ &$-$ & $\checkmark$ & $-$ \\
$0.31$ & $-$ & $-$ &$-$ & $-$     & $-$ \\
$0.5$  & $-$ & $-$ &$-$ & $\checkmark$ & $-$ \\ 
$0.7$  & $-$ & $-$ &$-$ & $\checkmark$ & $-$ \\
$0.8$  & $-$ & $-$ &$-$ & $-$     & $-$ \\
$1.0$  & $-$ & $-$ &$-$ & $\checkmark$ & $-$ \\
$1.44$ & $-$ & $-$ &$-$ & $-$     & $-$ \\
$1.5$  & $-$ & $-$ &$-$ & $\checkmark$ & $-$ \\
$1.67$ & $-$ & $-$ &$-$ & $-$     & $-$ \\
$2.0$  & $-$ & $-$ &$-$ & $\checkmark$ & $-$ \\
\hline
\end{tabular}
\end{center}
\begin{tablenotes}
\item In this Table the $\checkmark$ symbol indicates the actually exploited values for the initial galactocentric distance ($r_0$), eccentricity ($e$), and satellite mass ($M$) in the $N$-body simulations performed.
\end{tablenotes}
\label{RS1}
\end{table}

\begin{table}
\caption{}
\centering{Parameters defining the $N$-body $\gamma=1$ simulations.}
\begin{center}
\begin{tabular}{c c c c c c}
\hline
\hline
$r_0$ & $e=0$ & $e=0.50$ & \multicolumn{3}{c}{$e=1$} \\
\hline
$ $  &  $M=10^{-3}$ & $M=10^{-3}$ & $M=5\times 10^{-4}$ & $M=10^{-3}$ & $M=5\times 10^{-3}$ \\
\hline
$0.2$ & $\checkmark$ & $\checkmark$     &$\checkmark$ & $\checkmark$ & $\checkmark$ \\
$0.3$ & $\checkmark$ & $-$     & $-$     & $\checkmark$ & $\checkmark$ \\
$0.31$& $-$     & $-$     &$-$     & $\checkmark$ & $-$     \\
$0.5$ & $\checkmark$ & $\checkmark$     &$\checkmark$ & $\checkmark$ & $\checkmark$ \\ 
$0.7$ & $\checkmark$ & $-$     &$\checkmark$ & $\checkmark$ & $\checkmark$ \\
$0.8$ & $-$     & $-$     &$-$     & $\checkmark$ & $-$     \\
$1.0$ & $\checkmark$ & $\checkmark$ &$\checkmark$ & $\checkmark$ & $\checkmark$ \\
$1.44$& $-$     & $\checkmark$ &$-$     & $\checkmark$ & $-$     \\
$1.5$ & $-$     & $-$     &$-$     & $\checkmark$ & $\checkmark$ \\
$1.67$& $-$     & $-$     &$-$     & $\checkmark$ & $-$     \\
$2.0$ & $-$     & $-$     &$-$     & $\checkmark$ & $-$     \\
\hline
\end{tabular}
\end{center}
\begin{tablenotes}
\item All symbols as in Table \ref{RS1}.
\end{tablenotes}
\label{RS2}
\end{table}

\begin{table}
\caption{}
\centering{Parameters defining the $N$-body $\gamma=3/2$ simulations.}
\begin{center}
\begin{tabular}{c c c c c c}
\hline
\hline
$r_0$ & $e=0$ & $e=0.50$ & \multicolumn{3}{c}{$e=1$} \\
\hline
$ $  &  $M=10^{-3}$ & $M=10^{-3}$ & $M=5\times 10^{-4}$ & $M=10^{-3}$ & $M=5\times 10^{-3}$ \\
\hline
$0.2$  & $\checkmark$ & $-$ &$-$     & $\checkmark$ & $-$ \\
$0.3$  & $\checkmark$ & $-$ &$-$     & $\checkmark$ & $-$ \\
$0.31$ & $-$ & $-$ &$-$     & $-$     & $-$ \\
$0.5$  & $\checkmark$ & $-$ &$\checkmark$ & $\checkmark$ & $-$ \\ 
$0.7$  & $-$ & $-$ &$-$     & $\checkmark$ & $-$ \\
$0.8$  & $-$ & $-$ &$-$     & $-$     & $-$ \\
$1.0$  & $-$ & $-$ &$-$     & $\checkmark$ & $-$ \\
$1.44$ & $-$ & $-$ &$-$     & $-$     & $-$ \\
$1.5$  & $-$ & $-$ &$-$     & $-$     & $-$ \\
$1.67$ & $-$ & $-$ &$-$     & $-$     & $-$ \\
$2.0$  & $-$ & $-$ &$-$     & $\checkmark$ & $-$ \\
\hline
\end{tabular}
\end{center}
\begin{tablenotes}
\item All symbols as in Table \ref{RS1}.
\end{tablenotes}
\label{RS3}
\end{table}

\begin{table}
\caption{}
\centering{Fraction to the total of satellites sunk to the galactic center.}
\begin{center}
\begin{tabular}{ccccccccccccc}
\hline
\hline
\multicolumn{13}{c}{$M=10^{-6}$}\\
\hline
 & &\multicolumn{3}{c}{$t=500$ Myr}& &\multicolumn{3}{c}{$t=1$ Gyr}& &\multicolumn{3}{c}{$t=13.7$ Gyr}\\
\hline
$e$&&$r_{max}$&$f_0$&$f_{Pl}$& &$r_{max}$&$f_0$&$f_{Pl}$& &$r_{max}$&$f_0$&$f_{Pl}$\\
\hline
$0$  &&$0.342$&$0.017$&$0.030$& &$0.507$&$0.037$&$0.090$& &$2.293$&$0.335$&$0.765$\\
$1/2$&&$0.457$&$0.034$&$0.079$& &$0.677$&$0.065$&$0.170$& &$3.065$&$0.429$&$0.860$\\
$1$  &&$0.849$&$0.096$&$0.272$& &$1.258$&$0.175$&$0.487$& &$5.693$&$0.614$&$0.955$\\
\hline
\multicolumn{13}{c}{$M=10^{-4}$}\\
\hline
$0$  &&$1.92$&$0.284$&$0.699$& &$2.852$&$0.410$&$0.842$& &$12.90$&$0.799$&$0.991$\\
$1/2$&&$2.57$&$0.374$&$0.810$& &$3.809$&$0.500$&$0.904$& &$17.23$&$0.844$&$0.995$\\
$1$  &&$4.78$&$0.564$&$0.937$& &$7.077$&$0.672$&$0.970$& &$32.02$&$0.912$&$0.999$\\
\hline
\end{tabular}
\end{center}
\begin{tablenotes}
\small
\item The galaxy mass is assumed $M_G=10^{11}$M$_\odot$ and its length scale $a=250$ pc. 
      The galaxy density profile has $\gamma=0$. The fractions to the total satellite population decayed is $f_0$, assuming initial satellite distribution as a $\gamma=0$ profile, or $f_{Pl}$, assuming a Plummer profile.
\end{tablenotes}
\label{frac0}
\end{table}

\begin{table}
\caption{}
\centering{Fraction to the total of satellites sunk to the galactic center.}
\begin{center}
\begin{tabular}{c c c c c c c c c c c c c c}
\hline
\hline
\multicolumn{13}{c}{$M=10^{-6}$}\\
\hline
& &\multicolumn{3}{c}{$t=500$ Myr}& &\multicolumn{3}{c}{$t=1$ Gyr}& &\multicolumn{3}{c}{$t=13.7$ Gyr}\\
\hline
$e$&&$r_{max}$&$f_{3/2}$&$f_{Pl}$& &$r_{max}$&$f_{3/2}$&$f_{Pl}$& &$r_{max}$&$f_{3/2}$&$f_{Pl}$\\
\hline
$0$ & &$0.754$&$0.275$&$0.216$& &$1.12$&$0.385$&$0.417$& &$5.05$ &$0.763$&$0.944$\\
$1/2$&&$1.01$ &$0.357$&$0.354$& &$1.49$&$0.463$&$0.570$& &$6.77$ &$0.812$&$0.968$\\
$1$  &&$1.87$ &$0.525$&$0.390$& &$2.77$&$0.631$&$0.832$& &$12.55$&$0.891$&$0.999$\\
\hline
\multicolumn{13}{c}{$M=10^{-4}$}\\
\hline
$0$  &&$4.24$&$0.726$&$0.921$ & &$6.28$ &$0.801$&$0.963$& &$28.42$&$0.949$&$0.998$\\
$1/2$&&$5.66$&$0.783$&$0.944$ & &$8.39$ &$0.844$&$0.979$& &$38.10$&$0.962$&$0.999$\\
$1$  &&$10.52$&$0.872$&$0.987$& &$15.60$&$0.911$&$0.994$& &$70.55$&$0.979$&$0.999$\\
\hline
\end{tabular}
\end{center}
\begin{tablenotes}
\small
\item The galaxy mass is $M_G=10^{11}$M$_\odot$ and its length scale $a=250$ pc. 
      The galaxy density profile has $\gamma=3/2$. The fractions to the total satellite population decayed is $f_{3/2}$, assuming initial satellite distribution as a $\gamma=3/2$ profile, or $f_{Pl}$, assuming a Plummer profile. 

\end{tablenotes}
\label{frac3/2}
\end{table}

\begin{table}
\caption{}
\centering{Values of the parameters of the distribution functions of Eq. \ref{distribution function}}
\begin{center}
\begin{tabular}{c c c c c c}
\hline
\hline
\!\!\!$\gamma$\!\!\! & $0$ & $1$   & $4/3$      & $3/2$               & $7/4$ \\ 
\hline
\!\!\!$A_\gamma$ \!\!\!\! & $(3/2\pi^3)$ \!\!\!\!\! &$(8\sqrt 2\pi^3)^{-1}$ \!\!\!\!\! & \!\!\!\! $(62208\pi^3)^{-1}$ \!\!\!\! &  \!\!\!$3(32\sqrt 2\pi^3)^{-1}$\!\!\!\! &  \!\!\!$(1281280\sqrt 2\pi^3)^{-1}$\!\!\! \\
\!\!\!$B_\gamma$ \!\!\! & $0$ & $1$ & $-2\sqrt 2$\!\! & $8$ \!\! & $1$\\
\!\!\!$b_{0}$ \!\!\!  & $0$ & $3$   & $-54675$  \! \!& $-9/16$    \!\! & $-4188784600$\!\! \\
\!\!\!$b_{1}$ \!\!\!  & $0$ & $2$   & $186300$  \!\! & $-99/16$ \!\!   & $34508145672$ \!\!\\
\!\!\!$b_{2}$ \!\!\!  & $0$ & $-24$ & $-293328$ \!\! & $405/8$    \!\! & $-55318781804$\!\!\\
\!\!\!$b_{3}$ \!\!\!  & $0$ & $16$  & $206496$  \!\! & $-3705/56$ \!\! & $48778694536$ \!\!\\
\!\!\!$b_{4}$ \!\!\!  & $0$ & $0$   & $-67584$  \!\! & $561/14$   \!\! & $-28754568388$\!\!\\
\!\!\!$b_{5}$ \!\!\!  & $0$ & $0$   & $8192$   \!\!  & $-181/14$ \!\!  & $12242267940$ \!\!\\
\!\!\!$b_{6}$ \!\!\!  & $0$ & $0$   & $0$       \!\! & $15/7$     \!\! & $-3910165630$ \!\!\\
\!\!\!$b_{7}$ \!\!\!  & $0$ & $0$   & $0$      \! \! & $-1/7$   \!\!   & $697897200$   \!\!\\     
\!\!\!$b_{8}$ \!\!\!  & $0$ & $0$   & $0$      \!\!  & $0$      \!\!   & $955019800$  \!\! \\
\!\!\!$b_{9}$ \!\!\!  & $0$ & $0$   & $0$      \!\!  & $0$     \!\!    & $-179608380$ \!\! \\
\!\!\!$b_{10}$\!\!\!  & $0$ & $0$   & $0$      \!\!  & $0$     \!\!    & $25921460$   \!\! \\
\!\!\!$b_{11}$\!\!\!  & $0$ & $0$   & $0$      \!\!  & $0$     \!\!    & $-2828990$   \!\! \\
\!\!\!$b_{12}$\!\!\!  & $0$ & $0$   & $0$      \!\!  & $0$     \!\!    & $226548$     \!\! \\
\!\!\!$b_{13}$\!\!\!  & $0$ & $0$   & $0$      \!\!  & $0$     \!\!    & $-12586$     \!\! \\
\!\!\!$b_{14}$\!\!\!  & $0$ & $0$   & $0$      \! \! & $0$     \!\!    & $434$        \!\! \\
\!\!\!$b_{15}$\!\!\!\!  & $0$ & $0$   & $0$      \! \! & $0$     \!\!    & $-7$         \!\! \\
\hline
\end{tabular}
\end{center}
\end{table}

\end{document}